\documentclass[]{aa}
\pdfoutput=1

\usepackage{color}
\usepackage{graphicx}
\usepackage{txfonts}
\usepackage{lmodern}
\usepackage[colorlinks=true,linkcolor=blue,citecolor=blue,urlcolor=blue,bookmarks=false]{hyperref}
\usepackage{natbib,twoopt}
\usepackage{epstopdf}
\usepackage{lscape}

\bibpunct{(}{)}{;}{a}{}{,} %% natbib format for A&A and ApJ
\makeatletter
\newcommandtwoopt{\citeads}[3][][]{\href{http://adsabs.harvard.edu/abs/#3}%
{\def\hyper@linkstart##1##2{}%
\let\hyper@linkend\@empty\citealp[#1][#2]{#3}}}
\newcommandtwoopt{\citepads}[3][][]{\href{http://adsabs.harvard.edu/abs/#3}%
{\def\hyper@linkstart##1##2{}%
\let\hyper@linkend\@empty\citep[#1][#2]{#3}}}
\newcommandtwoopt{\citetads}[3][][]{\href{http://adsabs.harvard.edu/abs/#3}%
{\def\hyper@linkstart##1##2{}%
\let\hyper@linkend\@empty\citet[#1][#2]{#3}}}
\newcommandtwoopt{\citeyearads}[3][][]%
{\href{http://adsabs.harvard.edu/abs/#3}
{\def\hyper@linkstart##1##2{}%
\let\hyper@linkend\@empty\citeyear[#1][#2]{#3}}}
\makeatother

\begin{document}

\title{Properties of the CO and H$_2$O MOLsphere of the red supergiant \object{Betelgeuse} from VLTI/AMBER observations\thanks{Based on AMBER observations made with ESO Telescopes at the Paranal Observatory under programmes  ID 086.D-0351 and 286.D-5036(A)}}
\titlerunning{Properties of the MOLsphere of Betelgeuse from VLTI/AMBER observations}

   \author{M.~Montarg\`es\inst{1}
   			\and
          P.~Kervella\inst{1}
          \and
          G.~Perrin\inst{1}
          \and
          K.~Ohnaka\inst{2}
          \and
          A.~Chiavassa\inst{3}
          \and    
          S.~T.~Ridgway\inst{4}
          \and
          S.~Lacour\inst{1}
          }

   \institute{LESIA, Observatoire de Paris, CNRS, UPMC, Universit\'e Paris-Diderot, 5 place Jules Janssen, 92195 Meudon, France\\
              \email{miguel.montarges@obspm.fr}
         \and
         	Max-Planck-Institut f\"ur Radioastronomie, Auf dem H\"ugel 69, 53121 Bonn, Germany
         \and
         	Laboratoire Lagrange, UMR 7293, Universit\'e de Nice Sophia-Antipolis, CNRS, Observatoire de la C\^ote d'Azur, BP.~4229, 06304 Nice Cedex 4, France
         \and
         	National Optical Astronomy Observatory, PO Box 26732, Tuscon, AZ 85726, USA
             }
             
    \date{Received 29 January 2014 / Accepted 12 August 2014}

\abstract
	%Context heading
	{Betelgeuse is the closest red supergiant (RSG); therefore, it is well suited for studying the complex processes in its atmosphere that lead to the chemical enrichment of the interstellar medium.}
	%Aims heading
	{We intend to investigate the shape and composition of the close molecular layer (also known as the MOLsphere) that surrounds the star. This analysis is part of a wider program that aims at understanding the dynamics of the circumstellar envelope of Betelgeuse.}
	%Methods heading
	{On January and February 2011, Betelgeuse was observed using the Astronomical Multi-BEam combineR (AMBER) instrument of the Very Large Telescope Interferometer (VLTI) in the H and K bands. Using the medium spectral resolution of the instrument (R$\sim$1500), we were able to investigate the carbon monoxide band heads and the water-vapor bands. We used two different approaches to analyse our data: a model fit in both the continuum and absorption lines and then a fit with a Radiative HydroDynamics (RHD) simulation.}
	%Results heading
	{Using the continuum data, we derive a uniform disk diameter of $41.01 \pm 0.41$~mas, a power law type limb-darkened disk diameter of $42.28 \pm 0.43$~mas and a limb-darkening exponent of $0.155 \pm 0.009$. Within the absorption lines, using a single layer model, we obtain parameters of the MOLsphere. Using a RHD simulation, we unveil the convection pattern in the visibilities.}
	%Conclusions heading
	{We derived a new value of the angular diameter of Betelgeuse in the K band continuum. Our observations in the absorption lines are well reproduced by a molecular layer at 1.2 stellar radii containing both CO and H$_2$O. The visibilities at higher spatial frequencies are matching a convection pattern in a RHD simulation.}


\maketitle

%_______________________________

\section{Introduction}\label{Sect_Introduction}
	
	Betelgeuse ($\alpha$ Ori, HD 39801, HR 2061) is an M2Iab star, a prototype for the cool red supergiant class. These kind of stars are the expected progenitors of type IIP supernova, the most common kind of core-collapse supernova, and participate in the chemical enrichment of the interstellar medium (ISM), as they experience intensive mass loss. This process is not yet understood well and is essential to model the evolution of those stars.
	
	Being the closest red supergiant, Betelgeuse exhibits a very high brightness and a large apparent diameter. It was the first star (except for the Sun) to have its diameter measured \citepads{1921ApJ....53..249M} with the stellar interferometer at Mount Wilson Observatory. Since then, various observations were performed on Betelgeuse to study its circumstellar environment (CSE).  \citetads{2000ApJ...538..801T} proposed a non-photospheric molecular layer (the MOLsphere) with an effective temperature of $1500 \pm 500$~K and an H$_2$O column density of $10^{20}$~cm$^{-2}$ to fit their spectroscopic observations of Betelgeuse. Similar values were obtained by \citetads{2004A&A...421.1149O} and \citetads{2006ApJ...645.1448T}. \citetads{2004A&A...418..675P} derived parameters for both the photosphere and the MOLsphere using the IOTA interferometer: they obtained temperatures (T$_\mathrm{eff}^\mathrm{phot} = 3641 \pm 53$~K and T$_\mathrm{eff}^\mathrm{mol} = 2055 \pm 25$~K) and sizes ($\theta_\star = 43.76 \pm 0.12$~mas and $R_\mathrm{MOL} = 1.33$~R$_\star$) as well as the optical thickness ($\tau_\mathrm{K} = 0.060 \pm 0.003$, $\tau_\mathrm{L} = 0.026 \pm 0.002$ and $\tau_{11.5~\mu\mathrm{m}} = 2.33 \pm 0.23$). Its composition was explored by \citetads{2007A&A...474..599P}, who found evidence of the presence of H$_2$O and SiO using the MID-infrared Interferometric instrument (MIDI) of the Very Large Telescope Interferometer (VLTI) in the N band; they also derived the column density of the dust species Al$_2$O$_3$.
	
	Complex processes are ongoing, as material is moving away from the star, cooling and becoming chemically more complex. \citetads{2009A&A...504..115K} observed an asymmetric gas shell extending up to 6 stellar radii with adaptive optics observations between 1.04 and 2.17~$\mu$m using the Nasmyth Adaptive Optics System and the COnica detector (NACO) at the Very Large Telescope (VLT) and a dust shell further away (up to 2 to 3 arcsec) according to images obtained with the VLT Imager and Spectrometer for mid InfraRed (VISIR) instrument between 7.76 and 19.50 $\mu$m \citepads{2011A&A...531A.117K}. This envelope shows significant inhomogeneities and various structures, which suggest an asymmetric mass loss from the star that may continuously or episodically  occur \citepads{1996ApJ...463..336B}. Recently, \citetads{2013MNRAS.432L..61R} observed hot spots around Betelgeuse at $\sim 5 R_\star$ with an arc of 0.2-0.3 arcsec to the southwest using e-MERLIN (the upgrade of the Multi-Element Radio Linked Interferometer Network, MERLIN) at 5.5-6.0 GHz. 
	
	With the Astronomical Multi-BEam combineR (AMBER) of the VLTI, \citetads{2011A&A...529A.163O} spatially resolved upwelling and downdrafting gas motions within 1.5 R$_\star$ by exploring the red and blue wings of the first CO overtone lines, which was a much needed step in observing and understanding the dynamics of the stellar atmosphere and envelope.
	
	One of the first image reconstruction attempts by \citetads{1985ApJ...295L..21R} with the Canada France Hawaii Telescope (CFHT) in the visible domain revealed asymmetries in the star envelope. Closer to the star, \citetads{2009A&A...508..923H} reconstructed a high dynamic range image of the photosphere in the H band showing inhomogeneities, particularly two bright spots that compare well with 3D hydrodynamical simulations of RGSs \citepads{2010A&A...515A..12C}. Another spot was observed by \citetads{1998AJ....116.2501U} in the hot chromosphere using the Hubble Space Telescope, which was apparently fixed considering velocity measurements. They proposed that this spot could coincide with the south pole of Betelgeuse.
	
	Each observed layer of the CSE seems to present inhomogeneous structures. However, the process that links each shell is still unclear and requires further studies. Our observations with VLTI/AMBER at medium spectral resolution allow us to investigate the composition of the envelope and the shape of the photosphere. We present the data reduction process, which is quite unusual due to the large apparent size of the star, its brightness, and the use of diaphragms in Sect. \ref{Sect_AMBER_Observations}; then we fit the data with classical models (Sect. \ref{Sect_Model_AMBER}), and compare it with a radiative-hydrodynamics (RHD) simulation (Sect. \ref{Sect_Simulations}).

\section{Observations and data reduction} \label{Sect_AMBER_Observations}

	\subsection{VLTI Observations}

		We observed Betelgeuse with the ESO Very Large Telescope Interferometer \citepads[VLTI, ][]{2010SPIE.7734E...3H} using the Astronomical Multi-BEam combineR, AMBER \citepads{2007A&A...464....1P}. By combining three telescopes in the J, H, and K band, AMBER gives us information about the object's Fourier transform. The instrument measures the visibilities, which are directly its amplitude; and the differential phases (DP) are linked to the photocenter shift in a spectral line compared to the continuum. The closure phase (CP) is also obtained. It is defined as the sum of the three phases along the closed triangle formed by the three baselines: $\phi_{CP} = \phi_{12} + \phi_{23} + \phi_{31}$. It is mostly independent from atmospheric perturbations. Visibilities give us information on the size and shape of the star, while non-zero or non-$\pi$ CP indicates asymmetries in the object.
	
\begin{figure}[!ht]
		\centering
		\resizebox{\hsize}{!}{\includegraphics{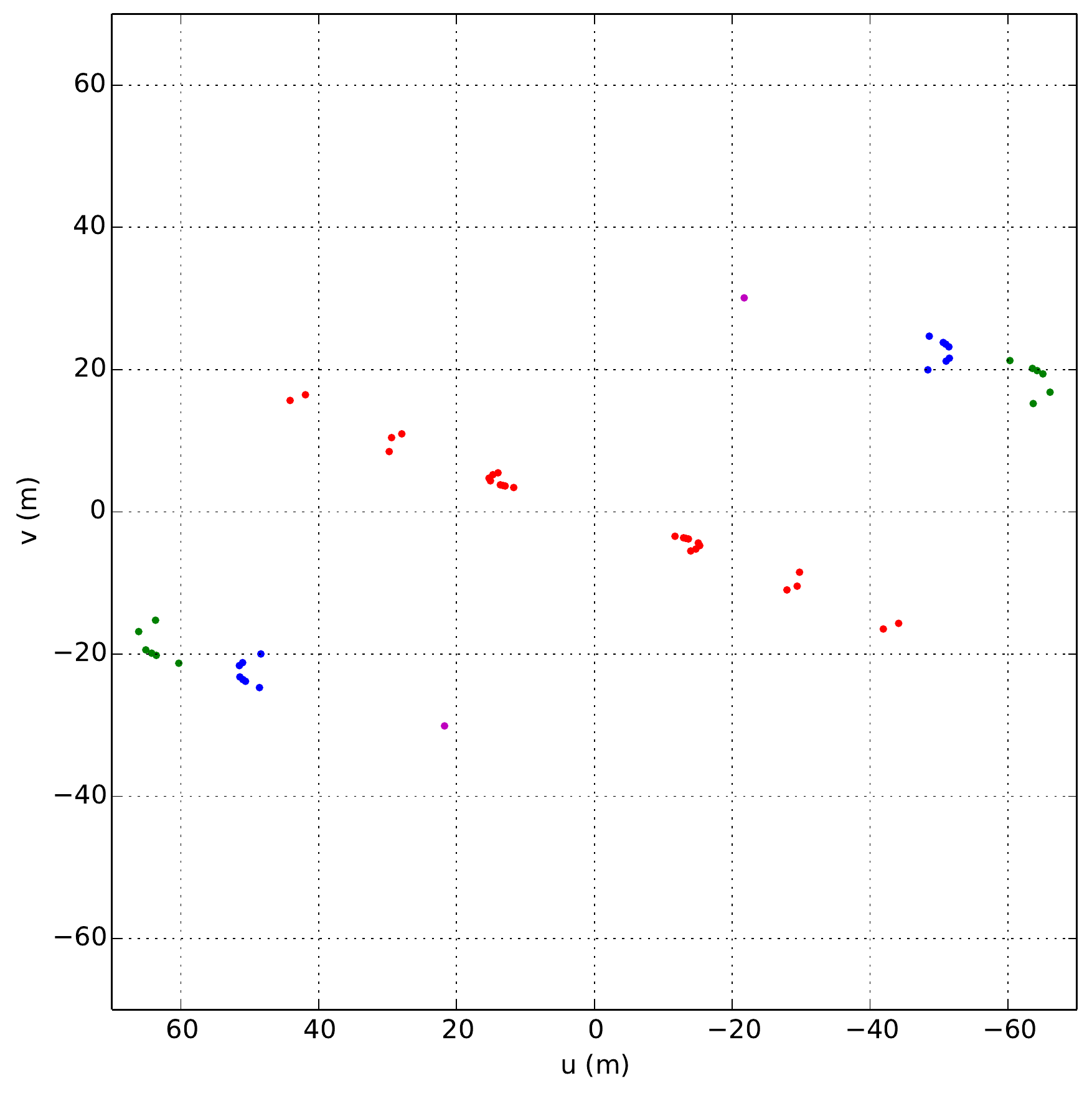}}
		\caption{(u,v) coverage of our AMBER data with a color-coded PA: red for PA $\sim 71^\circ$, green for $\sim 105^\circ$, blue for $\sim 115^\circ$, and violet for $\sim 145^\circ$. (North is 0$^\circ$ at the top and east is 90$^\circ$ on the left).}
	\label{Fig_uv_cov}
	\end{figure}

		The observations were performed on 2011 January 1, 2 and 3 and February 17 using three 1.8 m Auxiliary Telescopes (ATs) in the G0-H0-I1, E0-G0-I1, and E0-G0-H0 configurations. We used the medium spectral resolution mode ($R = \lambda / \Delta \lambda \sim 1500$) in the H and K bands (MR$\_$H 1.65 and MR$\_$K 2.3 instrument setups). The log of our AMBER observations is given in Table \ref{table_Obs}, and our (u,v) coverage is plotted in Fig. \ref{Fig_uv_cov}. The stars \object{HR 1543}, \object{HR 2275}, \object{HR 2469}, \object{HR 2508}, and \object{HR 3950} were observed as interferometric calibrators. The Fringe-tracking Instrument of NIce and TOrino (FINITO) was used in parallel with AMBER. As Betelgeuse is very bright, diaphragms were inserted in the beams to lower the incoming flux and avoid saturation of the detector. Data of 2011 January 1st are not used below as they were taken to obtain a suitable configuration of the instrument and are of poor quality.

%Electronic table (detail of observations)

\onllongtab{
%\begin{longtab}
\begin{landscape}
\begin{longtable}{lllllllllll} 
\caption{\label{table_Obs} Details of our AMBER observations of Betelgeuse. $B_p$ is the projected baseline length and PA is the position angle of the projected baseline vector ($0^\circ$ is north and $90^\circ$ is east.)}\\         
\hline\hline   
\# & Name & Mode & t$_{obs}$ & Stations & $B_\mathrm{p}$ & PA & Seeing (") & $\tau_0$ & DIT & Number\\
   &      &      &  (UTC)  &      &(m)& ($^\circ$) & (Visible) & (ms) & (ms) & of frames\\
\hline    
\endfirsthead  
\caption{continued.}\\   
\hline\hline                
\# & Name & Mode & t$_{obs}$ & Stations & $B_\mathrm{p}$ & PA & Seeing (") & $\tau_0$ & DIT & Number\\
   &      &      &  (UTC)  &      &(m)& ($^\circ$) & (Visible) & (ms) & (ms) & of frames\\
\hline                    
\endhead
\hline
\endfoot
\multicolumn{11}{c}{2011 January 01}\\
\hline
Cal & HR-1543 & MR\_K & 00:50 & E0-G0-H0 & 37.98/12.67/25.31 & 73.62/73.62/73.62 & 0.99 & 3.50 & 0.5 & 11*120\\
Cal & HR-2508 & MR\_K & 02:28 & E0-G0-H0 & 25.89/12.95/38.85 & 61.01/60.99/61.01 & 1.13 & 3.03 & 0.5 & 11*120\\
1 & Betelgeuse & MR\_K & 03:17 & E0-G0-H0 & 30.39/15.20/45.59 & 74.28/74.27/74.27 & 1.02 & 3.35 & 0.5 & 5*120\\
2 & Betelgeuse & MR\_K & 03:29 & E0-G0-H0 & 30.89/15.45/46.34 & 74.12/74.11/74.12 & 1.19 & 2.85 & 0.187 & 5*120\\
Cal & HR-2508 & MR\_K & 03:51 & E0-G0-H0 & 30.46/15.24/45.70 & 68.59/68.58/68.59 & 1.14 & 2.97 & 0.187 & 2*1000\\
Cal & HR-2508 & MR\_K & 04:03 & E0-G0-H0 & 30.99/15.51/46.50 & 69.55/69.53/69.55 & 1.08 & 3.14 & 0.187 & 10*100\\
11 & Betelgeuse & MR\_K & 04:21 & E0-G0-H0 & 31.95/15.99/47.94 & 73.01/73.00/73.01 & 1.10 & 3.09 & 0.187 & 10*100\\
Cal & HR-2508 & MR\_K & 04:35 & E0-G0-H0 & 31.75/15.89/47.64 & 71.28/71.26/71.28 & 0.83 & 4.06 & 0.187 & 10*100\\
4 & Betelgeuse & MR\_K & 04:51 & E0-G0-H0 & 31.90/15.96/47.86 & 72.03/72.01/72.02 & 0.77 & 4.40 & 0.187 & 10*100\\
Cal & HR-2508 & MR\_K & 05:06 & E0-G0-H0 & 31.99/16.01/48.00 & 72.63/72.61/72.62 & 0.75 & 4.48 & 0.187 & 10*100\\
5 & Betelgeuse & MR\_K & 05:28 & E0-G0-H0 & 31.19/15.61/46.79 & 70.40/70.39/70.40 & 0.90 & 3.74 & 0.187 & 3*200\\
6 & Betelgeuse & MR\_K & 05:33 & E0-G0-H0 & 31.02/15.52/46.54 & 70.12/70.11/70.12 & 0.89 & 3.77 & 0.187 & 9*100\\
Cal & HR-2469 & MR\_K & 05:51 & E0-G0-H0 & 31.23/15.63/46.86 & 74.28/74.27/74.28 & 0.84 & 3.84 & 0.187 & 11*100\\
Cal & HR-2469 & MR\_K & 06:35 & E0-G0-I1 & 58.57/48.00/14.75 & 114.06/125.21/75.10 & 0.78 & 4.12 & 0.187 & 8*100\\
Cal & HR-3950 & MR\_K & 07:21 & E0-G0-I1 & 68.79/56.25/15.19 & 105.10/113.00/74.47 & 1.16 & 2.73 & 0.187 & 10*100\\
Cal & HR-3950 & MR\_K & 08:14 & E0-G0-I1 & 66.95/53.79/15.93 & 103.62/112.20/73.38 & 0.91 & 3.42 & 0.187 & 22*100\\
\hline                    
\multicolumn{11}{c}{2011 January 02}\\
\hline
Cal & HR-1543 & MR\_K & 01:50 & E0-G0-I1 & 56.52/14.83/68.64 & 113.51/74.22/105.65 & 1.24 & 2.30 & 0.5 & 10*100\\
7 & Betelgeuse & MR\_K & 02:14 & E0-G0-I1 & 56.02/13.54/66.86 & 115.08/74.30/107.48 & 1.0 & 2.86 & 0.5 & 10*100\\
8 & Betelgeuse & MR\_K & 02:29 & E0-G0-I1 & 56.38/14.07/67.76 & 114.40/74.41/106.74 & 1.12 & 2.54 & 0.187 & 10*100\\
Cal & HR-2275 & MR\_K & 02:55 & E0-G0-I1 & 55.96/14.50/67.17 & 113.34/69.46/104.74 & 0.98 & 2.29 & 0.187 & 10*100\\
Cal & HR-2275 & MR\_K & 03:05 & E0-G0-I1 & 56.23/14.76/67.72 & 113.35/69.96/104.74 & 0.98 & 2.89 & 0.5 & 6*100\\
9 & Betelgeuse & MR\_K & 03:39 & E0-G0-I1 & 55.37/15.67/68.31 & 112.58/73.89/104.34 & 0.79 & 3.59 & 0.5 & 5*100\\
Cal & HR-2275 & MR\_K & 04:02 & E0-G0-I1 & 56.10/15.80/68.64 & 114.15/72.04/105.27 & 0.92 & 3.10 & 0.5 & 7*100\\
10 & Betelgeuse & MR\_K & 04:22 & E0-G0-I1 & 52.55/16.00/65.68 & 112.42/72.84/103.49 & 0.79 & 3.59 & 0.5 & 5*100\\
Cal & HR-2275 & MR\_K & 04:45 & E0-G0-I1 & 54.09/16.00/66.72 & 115.67/72.83/106.29 & 0.98 & 2.91 & 0.5 & 6*100\\
11 & Betelgeuse & MR\_K & 05:19 & E0-G0-H0 & 31.29/15.66/46.94 & 70.58/70.57/70.58 & 1.19 & 2.39 & 0.5 & 5*100\\
Cal & HR-2275 & MR\_K & 05:36 & E0-G0-H0 & 31.12/15.57/46.69 & 73.03/73.01/73.02 & 1.23 & 2.44 & 0.5 & 5*100\\
12 & Betelgeuse & MR\_K & 05:52 & E0-G0-H0 & 30.14/15.08/45.23 & 68.75/68.74/68.74 & 1.09 & 2.76 & 0.5 & 5*120\\
Cal & HR-2275 & MR\_K & 06:21 & E0-G0-H0 & 29.19/14.61/43.80 & 72.47/72.46/72.46 & 1.14 & 2.68 & 0.5 & 10*120\\
Cal & HR-3950 & MR\_K & 07:44 & E0-G0-H0 & 15.67/31.32/46.99 & 74.00/74.01/74.00 & 1.18 & 2.75 & 0.5 & 6*120\\
Cal & HR-2508 & MR\_K & 08:01 & E0-G0-H0 & 11.67/23.32/34.99 & 74.81/74.83/74.82 & 0.99 & 3.30 & 0.5 & 20*100\\
\hline
\multicolumn{11}{c}{2011 January 03}\\
\hline
Cal & HR-1543 & MR\_K & 01:03 & E0-G0-I1 & 66.73/55.95/13.51 & 107.44/115.08/74.01 & 0.99 & 4.02 & 0.5 & 5*100\\
Cal & HR-1543 & MR\_K & 01:17 & E0-G0-I1 & 67.58/56.31/13.99 & 106.79/114.49/74.16 & 0.86 & 4.62 & 0.5 & 5*100\\
13 & Betelgeuse & MR\_K & 01:33 & E0-G0-I1 & 63.68/54.36/12.09 & 109.60/117.10/73.61 & 0.86 & 4.58 & 0.5 & 5*100\\
Cal & HR-2508 & MR\_K & 01:49 & E0-G0-I1 & 55.95/48.40/11.82 & 102.08/112.10/56.6 & 0.85 & 4.61 & 0.5 & 5*100\\
14 & Betelgeuse & MR\_K & 02:05 & E0-G0-I1 & 66.51/55.86/13.36 & 107.73/115.32/74.24 & 0.63 & 6.24 & 0.5 & 5*100\\
Cal & HR-2275 & MR\_K & 02:19 & E0-G0-I1 & 64.49/54.43/13.48 & 104.93/113.61/67.38 & 0.70 & 5.63 & 0.5 & 5*120\\
15 & Betelgeuse & MR\_K & 02:33 & E0-G0-I1 & 68.12/56.49/14.31 & 106.39/114.10/74.43 & 0.63 & 6.17 & 0.5 & 5*120\\
Cal & HR-2508 & MR\_K & 03:03 & G0-H0-I1 & 40.77/54.79/28.65 & 142.77/112.12/65.62 & 0.63 & 6.15 & 0.5 & 6*100\\
16 & Betelgeuse & MR\_K & 03:20 & G0-H0-I1 & 37.24/55.95/30.85 & 143.99/112.81/74.14 & 0.69 & 5.61 & 0.5 & 5*120\\
Cal & HR-2275 & MR\_K & 03:34 & G0-H0-I1 & 39.57/56.51/30.91 & 145.38/113.65/71.34 & 0.65 & 6.00 & 0.5 & 5*120\\
17 & Betelgeuse & MR\_K & 03:50 & G0-H0-I1 & 35.91/54.59/31.66 &146.00/112.44/73.60  & 0.74 & 5.22 & 0.5 & 5*100\\
Cal & HR-2275 & MR\_K & 04:02 & G0-H0-I1 & 38.80/55.97/31.66 & 147.43/114.26/72.16 & 0.89 & 4.30 & 0.5 & 5*100\\
Cal & HR-2275 & MR\_H & 04:21 & G0-H0-I1 & 38.23/55.23/31.92 & 149.05/114.85/72.55 & 1.14 & 3.38 & 0.5 & 5*100\\
18 & Betelgeuse & MR\_H & 04:43 & G0-H0-I1 & 33.35/50.28/31.90 & 151.23/112.68/72.03 & 0.92 & 4.16 & 0.5 & 6*100 \\
Cal & HR-2508 & MR\_H & 05:14 & G0-H0-I1 & 38.73/54.48/31.90 & 152.74/117.59/73.25 & 0.98 & 3.90 & 0.5 & 5*100\\
Cal & HR-2508 & MR\_H & 05:39 & E0-G0-I1 & 15.76/52.81/64.82 & 73.99/119.54/109.55 & 0.72 & 4.52 & 0.5 & 5*120\\
19 & Betelgeuse & MR\_H & 05:57 & E0-G0-I1 & 14.89/40.67/51.90 & 68.16/115.71/103.49 & 0.93 & 3.47 & 0.5 & 10*100\\
Cal & HR-2508 & MR\_H & 06:22 & E0-G0-I1 & 15.01/49.01/59.94 & 74.89/123.86/112.97 & 1.05 & 3.07 & 0.5 & 5*100\\
Cal & HR-2508 & MR\_H & 06:41 & E0-G0-H0 & 43.57/14.53/29.04 & 75.14/75.12/75.14 & 0.78 & 4.14 & 0.5 & 5*100\\
Cal & HR-2508 & MR\_H & 07:32 & E0-G0-I1 & 48.84/41.12/12.78 & 121.77/134.82/75.19 & 0.58 & 5.58 & 0.5 & 10*100\\
Cal & HR-2508 & MR\_H & 08:09 & E0-G0-I1 & 42.03/36.77/11.12 & 129.39/143.71/74.51 & 0.72 & 4.45 & 0.5  & 15*100\\
\hline
\multicolumn{11}{c}{2011 February 17}\\
\hline
Cal & HR-2275 & MR\_H & 02:09 & E0-G0-H0 & 47.72/31.80/15.91 & 73.01/73.02/73.00 & 1.39 & 2.59 & 0.5 & 5*100\\
20 & Betelgeuse & MR\_H & 02:47 & E0-G0-H0 & 45.70/30.46/15.24 & 69.22/69.23/69.22 & 1.47 & 2.45 & 0.5 & 6*100\\
\end{longtable}
\end{landscape}
%\end{longtab}
}

	\subsection{Data reduction}\label{SubSect_Reduction}
	
		The data were reduced using the \texttt{AMBER data reduction package} version 3.0.3 also known as \texttt{amdlib}. The reduction package uses the P2VM algorithm described in \citetads{2007A&A...464...29T} and \citetads{2009A&A...502..705C}.	The procedure is straightforward, but the visibilities are unusually low and the error bars are underestimated as Betelgeuse has a large apparent diameter. We split each dataset into five subsets on which we performed the reduction process using \texttt{amdlib} and estimated new error bars for the visibilities, DPs and CPs, by computing the standard deviation within each subset \citepads{2009A&A...503..183O}.
		
		We checked the effects of the data frame selection on the signal-to-noise ratio (S/N) by using different criteria, but we did not observe significant changes on the observables. Thus, we decided to average the data keeping the best $80 \%$ of all frames.
		
		We used the telluric line template from NSO/Kitt Peak FTS produced by NSF/NOAO\footnote{\href{http://www.eso.org/sci/facilities/paranal/instruments/isaac/tools/spectroscopic$\_$standards.html$\#$Tellur}{http://www.eso.org/sci/facilities/paranal/instruments/} \href{http://www.eso.org/sci/facilities/paranal/instruments/isaac/tools/spectroscopic$\_$standards.html$\#$Tellur}{isaac/tools/spectroscopic$\_$standards.html$\#$Tellur}} to perform the wavelength calibration. We identified telluric absorption features along our spectral domain and fitted their wavelength with a quadratic law: 

		\begin{equation}		
			\lambda_\mathrm{calib} = a\lambda_\mathrm{RAW}^2 + b\lambda_\mathrm{RAW} + c.
		\end{equation}		
		
		\begin{table}[h!]
		\caption{Interferometric calibrators (angular diameters values from \citeads{2010SPIE.7734E.140L}).}             
		\label{table_Cal}      
		\centering          
		\begin{tabular}{c c c c c} 
		\hline\hline
		HR Identifier & Spectral type & \multicolumn{2}{c}{UD diameter (mas)}\\
		  & & H band  & K band &\\
		\hline
		HR 1543 &  F6V  & 2.160 & 2.166\\
		HR 2275 &  M1III & 3.653 & 3.682\\
		HR 2469 &  M0III & 2.499 & 2.519\\
		HR 2508 &  M1Iab & 5.050 & 4.370\\
		HR 3950 &  M2III & 4.574 & 4.610\\
		\hline
		\end{tabular}
		\end{table}		
		
\begin{figure}[!ht]
		\centering
		\resizebox{\hsize}{!}{\includegraphics{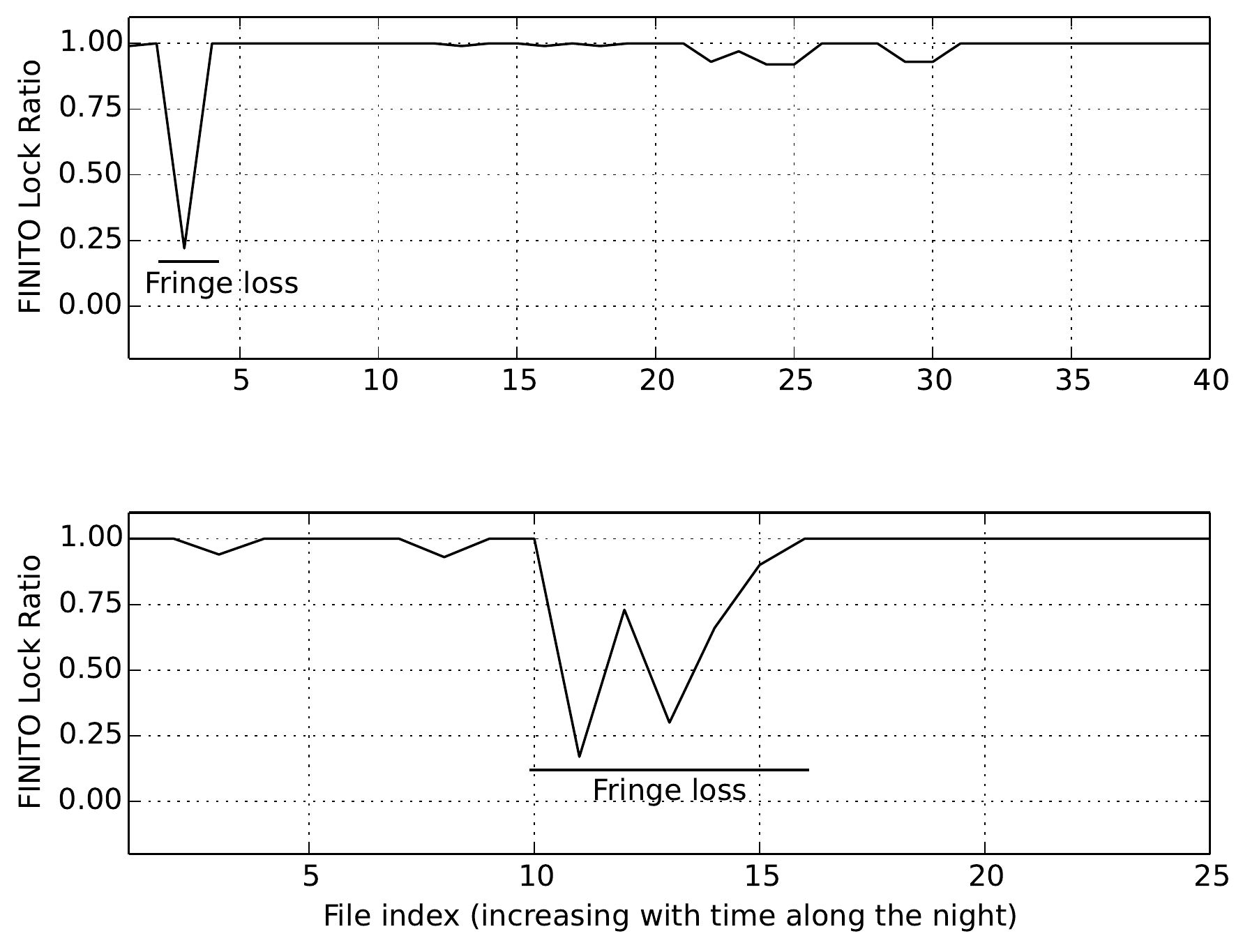}}
		\caption{FINITO lock ratio evolution: fraction of time with locked fringes during the exposure time. \textit{Top:} 2011-01-02 (K band).  \textit{Bottom:} 2011-01-03 (K band)}
		\label{Fig_FinitoRatio}
		\end{figure}

		We computed the interferometric transfer function for each night and configuration. The diameters assumed for the calibrators are compiled in Table \ref{table_Cal}. The transfer function was stable along the four nights but some isolated datasets showed low visibilities: our analysis revealed that the fringe tracker FINITO was occasionally losing the fringes and causing a strong decrease of the measured visibilities. As the real-time fringe tracker data were not yet available, we used the lock ratio keywords in the RAW file that quantify the fraction of time of fringe lock during each exposure to discriminate biased datasets (Fig. \ref{Fig_FinitoRatio}).
		
		The deepest data points in the CO band heads were showing inconsistent square visibilities ($\sim -10^{11}$) over a range from 1 to 4~nm, which is how \texttt{amdlib} tags low quality fringe fitting. We decided to ignore these data points. 
				
\section{Fit with analytical models} \label{Sect_Model_AMBER}

\subsection{K-band continuum data ($\lambda \leq 2.245~\mu$m): Uniform and limb-darkend disk diameters} \label{SubSect_Continuum}

\begin{figure*}[!ht]
		\centering
		\resizebox{\hsize}{!}{\includegraphics[width=\columnwidth]{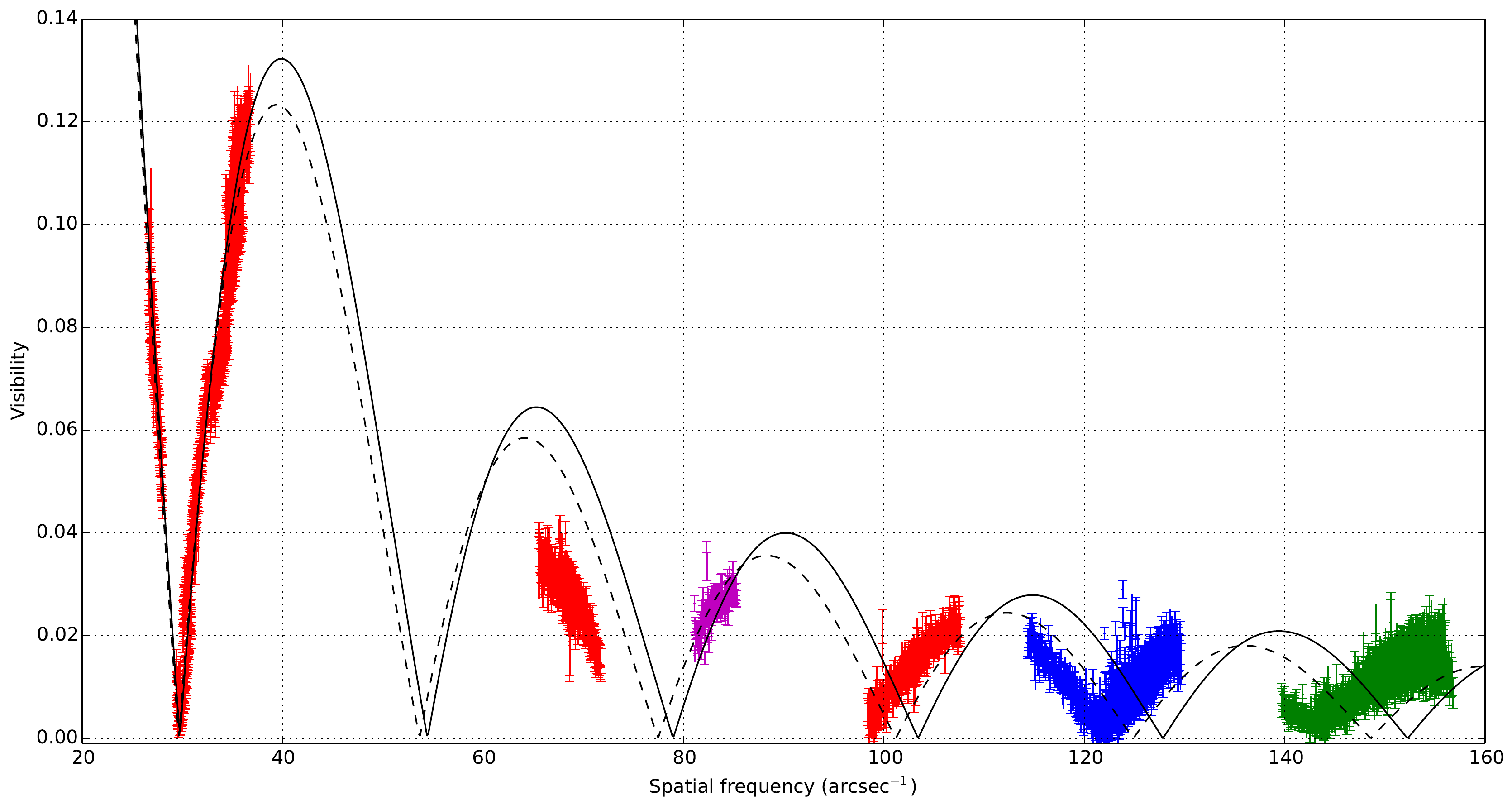}}
		\caption{Continuum visibilities with color-coded PA matching Fig. \ref{Fig_uv_cov}. The black continuous line represents the best fit UD model, and the black dashed line represents the best fit LDD model, whose results are presented in Table \ref{Table_Results_Continuum}.}
		\label{Fig_V2_PA}
	\end{figure*}

	Our K-band continuum data are composed of 59 datasets with 262 spectral channels in the continuum wavelength range. We fitted the data with a uniform disk (UD) and a power law type limb-darkened disk (LDD) described in \citetads{1997A&A...327..199H}. This LDD model introduces a second parameter, $\alpha$,  the exponent of the power law. We restrain ourselves to the low spatial frequencies (first and second lobe, spatial frequencies below 55 arcsec$^{-1}$) to avoid contamination by small scale structures. The results of these fits are presented in Table \ref{Table_Results_Continuum}. The best fit visibilities for each model are plotted in Fig. \ref{Fig_V2_PA} with the data.
	
	\begin{table}[h!]
		\caption{Best fit values for the uniform disk and limb-darkened disk models when considering all the observed PA and only the first and second lobes.}             
		\label{Table_Results_Continuum}      
		\centering          
		\begin{tabular}{l l l l} 
			\hline\hline
			Model & $\theta$ (mas) & $\alpha$ & $\chi^2$ \\
			\hline
			UD & $41.01 \pm 0.41$ & - & 5.27\\
			LDD & $42.28 \pm 0.43$ & $0.155 \pm 0.009$ & 4.89\\
			\hline
		\end{tabular}
	\end{table}

\begin{figure}[!ht]
		\centering
		\resizebox{\hsize}{!}{\includegraphics[viewport=39 110 571 520,clip]{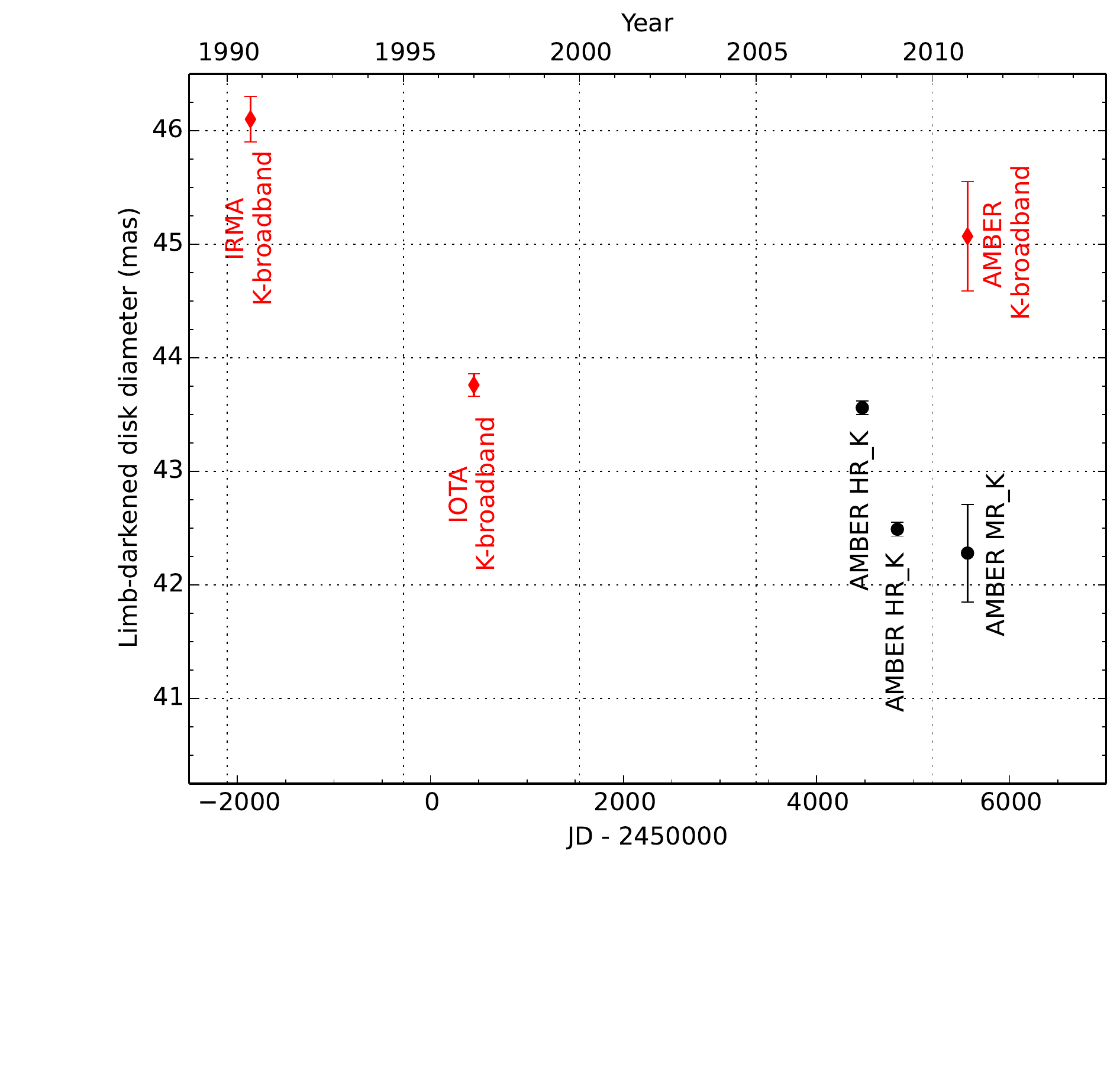}}
		\caption{Overview of limb-darkened disk measurements of Betelgeuse. The values considering only the continuum of the K band are displayed in black dots, and the K-broadband measurements are in red diamonds. The IRMA measurement comes from \citetads{1992AJ....104.1982D}, IOTA from \citetads{2004A&A...418..675P}, the two AMBER high resolution K-band measurements from \citetads{2009A&A...503..183O,2011A&A...529A.163O}, and the AMBER K-medium resolution and broadband are from this work.}
		\label{Fig_DiameterTime}
	\end{figure}

	Our UD diameters are lower than the previous measurements from \citetads{1992AJ....104.1982D} ($44.2 \pm 0.2$~mas) and \citetads{2004A&A...418..675P} ($43.26 \pm 0.04$~mas), but these values were obtained with K-broadband data and certainly contained contributions from the molecular opacities through the absorption features. However, our results agree with the UD diameter of \citetads{2011A&A...529A.163O} of $42.05 \pm 0.05$~mas and their LDD diameter value of $42.49 \pm 0.06$~mas. On Fig. \ref{Fig_DiameterTime}, we plotted previous measurements of the LDD diameter of Betelgeuse with time, again one can notice the greater apparent diameter obtained with K-broadband datasets: this is caused by the contamination by the molecular material around the star. It is also remarkable that even if the K-broadband and K-continuum diameters are not constant, they do not show a monotone variation.
	
	From our limb-darkened measurements and the distance of $197 \pm 45$~pc \citepads{2008AJ....135.1430H}, we derive the stellar radius of Betelgeuse $R_\star = 897 \pm 211~R_\odot$ and its luminosity $L_\star = 1.27 \pm 0,60 \times 10^5~L_\odot$ by considering an effective temperature of 3690~K \citepads{2011A&A...529A.163O}. As it has already been pointed out \citepads{2004A&A...418..675P}, the large uncertainty on the parallax of Betelgeuse is mainly responsible for the large error bars on these physical parameters.
	
	On Fig. \ref{Fig_V2_PA}, the observed visibilities in the continuum deviate strongly from the LDD model for spatial frequencies higher than 60 arcsec$^{-1}$. However, \citetads{2009A&A...503..183O,2011A&A...529A.163O} did not observe such deviations in their high spectral resolution dataset. Our dataset samples other directions of the (u,v) plane and not only the PA = 71.39$^\circ$ they covered. Figures \ref{Fig_uv_cov} and \ref{Fig_V2_PA} show the visibilities and the (u,v) plane with a color-coded PA. We fitted the main PA direction (71.39$^\circ$, in red) with UD and LDD models. The result of these fits are presented in Table \ref{Table_Results_PA}.
		
		\begin{table}[h!]
			\caption{Best fit values for the uniform disk and limb-darkened disk models for our main sampled (u,v) direction (PA = 71.39$^\circ$) and for all the spatial frequencies.}             
			\label{Table_Results_PA}      
			\centering          
			\begin{tabular}{l l l l} 
				\hline\hline
				Model & $\theta$ (mas) & $\alpha$ & $\chi^2$ \\
				\hline
				UD & $40.9 \pm 0.52$ & - & 110\\
				LDD & $41.8 \pm 0.57$ & $0.10 \pm 0.02$ & 55\\
				\hline
			\end{tabular}
		\end{table}
		
	Even for the main direction of PA = 71.39$^\circ$, we notice that we observe these deviations from the LDD. They can be explained in two different ways. First, there could be some residuals from the debiasing: the FINITO lock losses could still lower our visibilities, but they may not displace the zeros of the visibility function. This supports our second hypothesis: there has been a change on the photosphere of the star since \citetads{2009A&A...503..183O} observations, which makes it deviate from central symmetry. We develop this hypothesis in Sect. \ref{Sect_Simulations}.

\subsection{K-band, CO, and water absorption bands ($\lambda > 2.245~\mu$m)} \label{SubSect_CO}

	\subsubsection{Spherical thin layer: The MOLsphere}\label{SubSubSect_MOLsphere}

\begin{figure}[!ht]
		\centering
		\resizebox{\hsize}{!}{\includegraphics{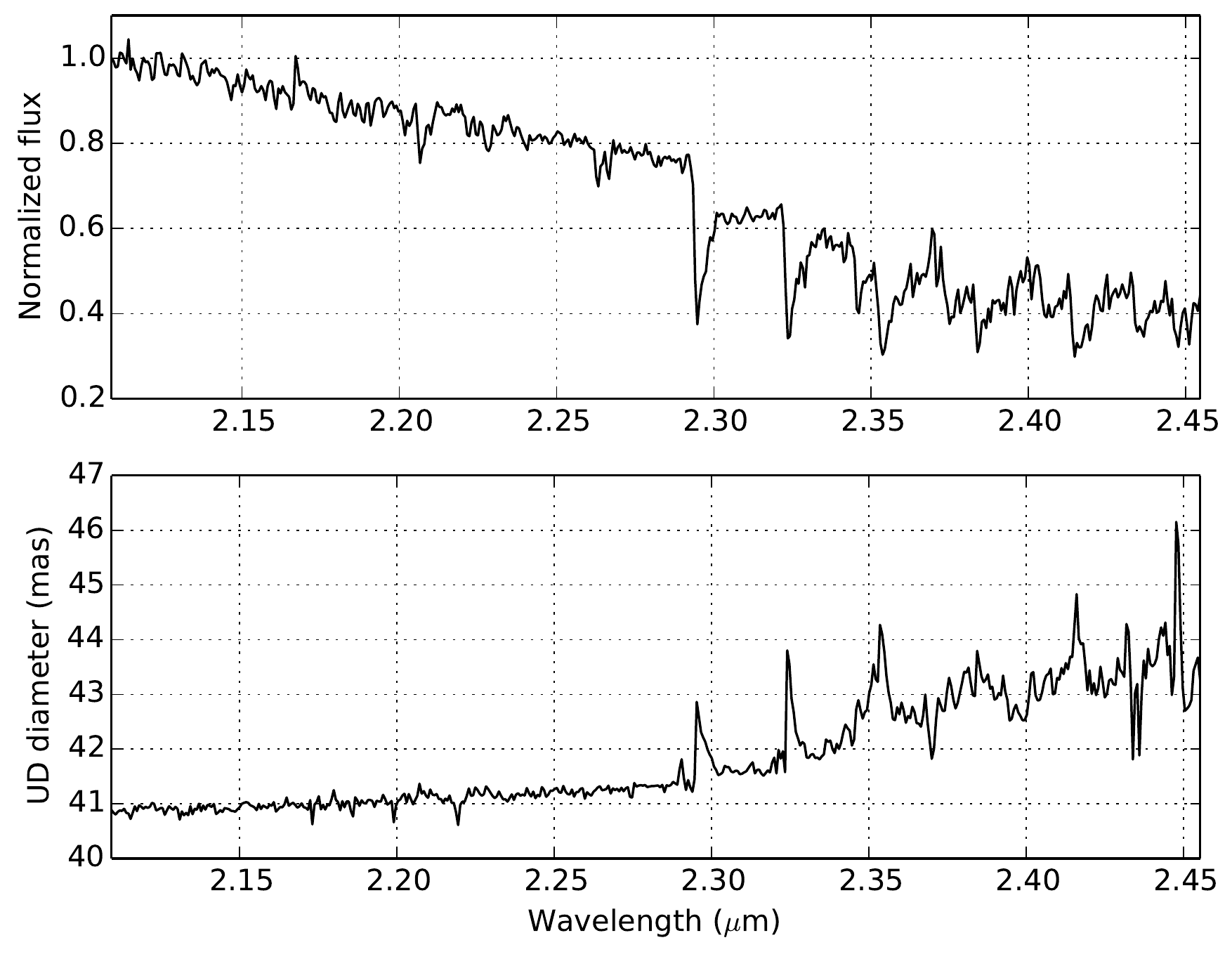}}
		\caption{\textit{Top:} Observed AMBER spectrum. \textit{Bottom:} Best fit UD diameter as a function of wavelength. The spectral channels with negative squared visibilities are ignored.}
	\label{Fig_UDAll}
	\end{figure}

	Figure \ref{Fig_UDAll} presents the result of the fit of the UD diameter as a function of wavelength, by considering only the first and second lobes of the visibility function. We used the whole spectral range of our AMBER data, except for the core of the CO absorption lines where we do not have visibility measurements. We had to ignore these spectral channels, and to interpolate over them (see Sect. \ref{SubSect_Reduction}). For wavelengths longer than $2.245~\mu$m, the UD diameter increases: this is the signature of the MOLsphere. The strong peaks are caused by CO as they clearly match the absorption lines in the spectrum.

	% 1 to 4 nm for bad vis

	\begin{figure} %[!ht]
		\centering
		\includegraphics[width=5.cm]{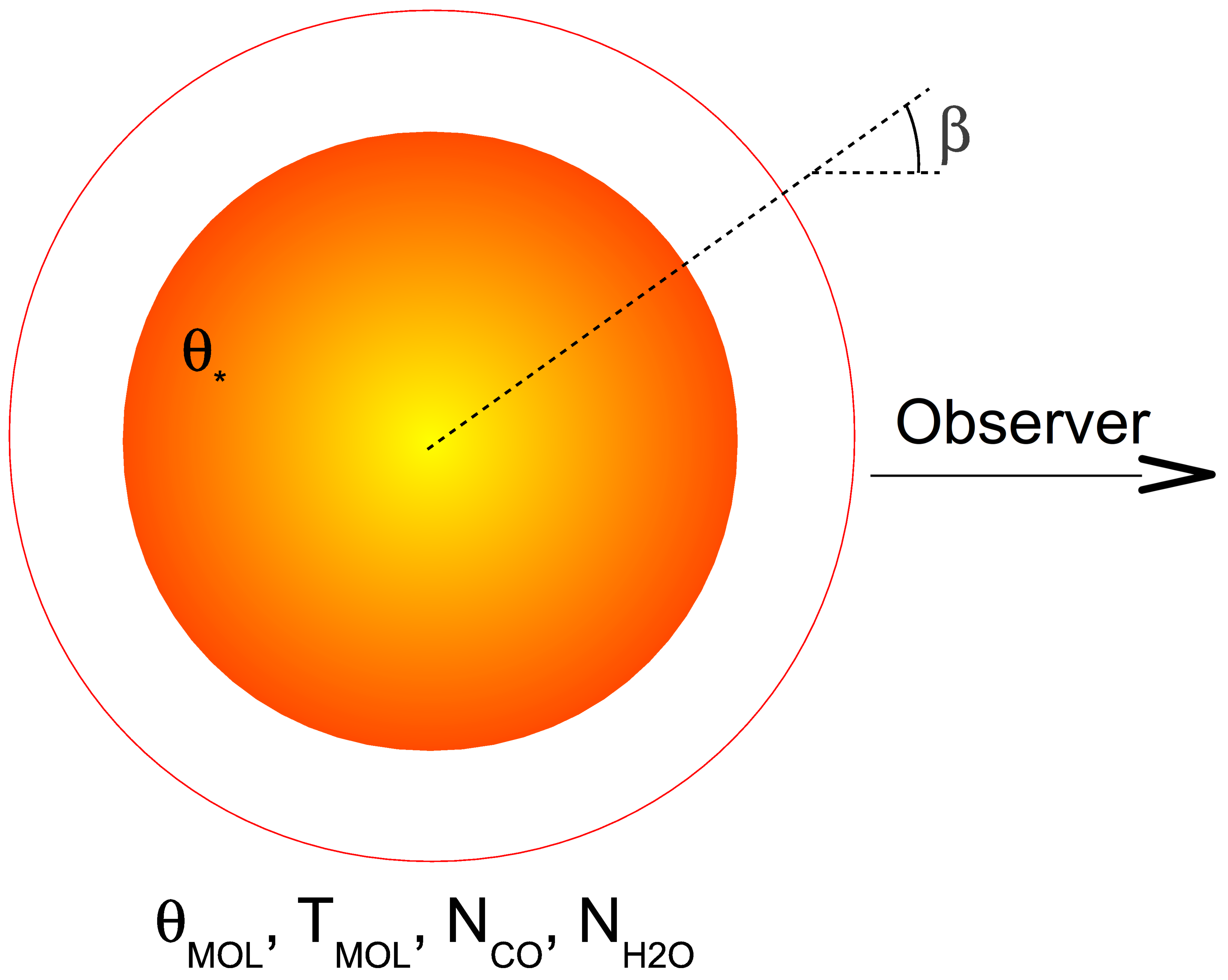}
		\caption{Illustration of the single layer model of the MOLsphere. The variable $\beta$ is the angle between the radius vector and the line of sight at the layer surface.}
		\label{Fig_ModelSchem}
	\end{figure}	
		
	To model this CO and H$_2$O envelope around Betelgeuse, we used a single thin layer model \citepads{2004A&A...418..675P}. We computed the opacity from the line list of \citetads{1994ApJS...95..535G} for CO, and of \citetads{1997JChPh.106.4618P} for H$_2$O; therefore, our parameter is not the optical depth $\tau$ but the column densities for both species. This MOLsphere surrounds a photosphere model computed from the Kurucz grid\footnote{\url{http://kurucz.harvard.edu/}} \citepads{2003IAUS..210P.A20C,2005MSAIS...8...14K} for T$_\mathrm{eff} = 3700~$K, $\log g = -0.5$, and solar metallicity. As our aim is to compute the column densities for both carbon monoxide and water vapor in the atmosphere of the star, we used the continuum fluxes given in the Kurucz model, which are free from any absorption lines. The layer absorbs the light from the star and re-emits it like a blackbody. We do not consider scattering at these wavelengths. The MOLsphere is assumed to be thin and at the local thermodynamic equilibrium (LTE). This star and thin layer model is illustrated on Fig. \ref{Fig_ModelSchem}. If $\sin (\beta) \leq \frac{\theta_\star}{\theta_{\mathrm{MOL}}}$ its analytical expression is given by 
	
	\begin{equation}
		\begin{array}{l}
			I_{N_{\mathrm{CO}},N_{{\mathrm{H}_2\mathrm{O}}}}(\lambda,\beta) \\
			 = I_{\mathrm{Kurucz}}(\lambda) \ \exp\left(\frac{-\tau(N_{\mathrm{CO}},N_{{\mathrm{H}_2\mathrm{O}}};\lambda)}{\cos(\beta)}\right)\\ 
			+ B(\lambda,T_{\mathrm{MOL}})\left[1-\exp\left(\frac{-\tau(N_{\mathrm{CO}},N_{{\mathrm{H}_2\mathrm{O}}};\lambda)}{\cos(\beta)}\right)\right].
		\end{array}
	\end{equation}
	
	If $\sin(\beta)\geq \frac{\theta_\star}{\theta_{\mathrm{MOL}}}$, then
	 
	\begin{equation}
		\begin{array}{l}
			I_{N_{\mathrm{CO}},N_{{\mathrm{H}_2\mathrm{O}}}}(\lambda,\beta)\\
			= B(\lambda,T_{\mathrm{MOL}})\left[1-\exp\left(\frac{-2\tau(N_{\mathrm{CO}},N_{{\mathrm{H}_2\mathrm{O}}};\lambda)}{\cos(\beta)}\right)\right].\\
		\end{array}
	\end{equation}
	
	This model is not physically accurate as CO in particular is continuously distributed from the photosphere to large distances from the star (with a continuous distribution of temperature and density), but it allows here  to get the typical characteristics of the MOLsphere in the field of view of the interferometer, which is close to the star. As the depths of the strong and populated low excitation bands of CO are dominated by the MOLsphere, our fit to those bands reflects the conditions in that component of the atmosphere. The five parameters of the model are the photospheric diameter $\theta_\star$, the MOLsphere diameter $\theta_{\mathrm{MOL}}$, the MOLsphere temperature $T_{\mathrm{MOL}}$, the CO, and the H$_2$O column densities $N_{\mathrm{CO}}$ and $N_{{\mathrm{H}_2\mathrm{O}}}$. The function $B(\lambda,T)$ is the Planck function; $\beta$ is the angle between the line of sight and the center of the star at the layer surface, and $\tau(N_{\mathrm{CO}},N_{{\mathrm{H}_2\mathrm{O}}};\lambda)$ is the MOLsphere optical depth computed from the previously indicated line lists.\\
	 
	 Then we computed the Hankel transform to get the visibility:				
		\begin{equation}
			V_\lambda (x) = \frac{ \int_0^1 I(\lambda,r)J_0(rx)rdr}{\int_0^1 I(\lambda,r)rdr}.
		\end{equation}
		
	With $x = {\pi B_p \theta_\star / \lambda}$, $r = \sin(\beta)$, $B_p$ is the projected baseline (see Table \ref{table_Obs}), and $J_0$ is the zeroth order Bessel function of the first kind.\\
	
	The deepest data points in the CO band heads were not usable because the data reduction package did not manage to recover the visibility from the fringes at those wavelengths (see Sect. \ref{SubSect_Reduction}). To compensate for this lack of information, we added the photometric spectrum provided by the AMBER instrument to our set of constraints. We only considered data in the first and second lobes of the visibility function to avoid contamination by small scale structures. To get the best fit parameters of our data, we minimized the $\chi^2$:
	
	\begin{equation}
		\begin{array}{l}
      	\chi^2(T_{\mathrm{MOL}},\theta_{\mathrm{MOL}})\\
      	= \sum_{i=1}^N \left(\frac{Y_i - M(T_{\mathrm{MOL}},\theta_{\mathrm{MOL}}, N_{\mathrm{CO}}, N_{{\mathrm{H}_2\mathrm{O}}};S_i)}{\sigma_i}\right)^2. \\
   		\end{array}  
   	\end{equation} 	
   	
	Here, $S_i = B_p / \lambda$ are the sampled spatial frequencies, $Y_i$ the AMBER data (spectrum and visibilities) in the absorption lines, and $M$ the corresponding value of the model.
   	
   	The five parameters of the model are not completely independent: correlations exist particularly between $T_\mathrm{MOL}$ and $\theta_\mathrm{MOL}$ on one hand and between the two column densities on the other hand. Our strategy to perform this model fitting was to constrain $\theta_\star$ to the best fit value of the UD diameter from continuum data (Sect. \ref{SubSect_Continuum}). Then, on a grid of $(T_\mathrm{MOL},\theta_\mathrm{MOL})$, we fitted one of the two column densities while keeping the other one constant. This gives us a column density map and a $\chi^2$ map (see examples of the $\chi^2$ map for both column densities on Fig. \ref{Fig_Chi2Map} online); we select our best fit column density value corresponding to the minimum $\chi^2$ and use it as input parameter to compute the maps of the second column density. By iterating the process for $N_\mathrm{CO}$ and $N_{\mathrm{H}_2\mathrm{O}}$ until the best fit values stay in their statistical error bars, we derived the best fit values for the MOLsphere parameters.	
   	
   	To avoid using initial conditions in this fitting process, we decided to initially only fit the CO column density. Indeed, this molecule has the strongest absorption lines. Therefore, we set the water vapor column density to zero in the first iteration. This allows to derive a first estimation of N$_\mathrm{CO}$, which is used as input for the N$_{\mathrm{H}_2\mathrm{O}}$ fit.
	
	\begin{figure}[!ht]
		\centering
		\resizebox{\hsize}{!}{\includegraphics{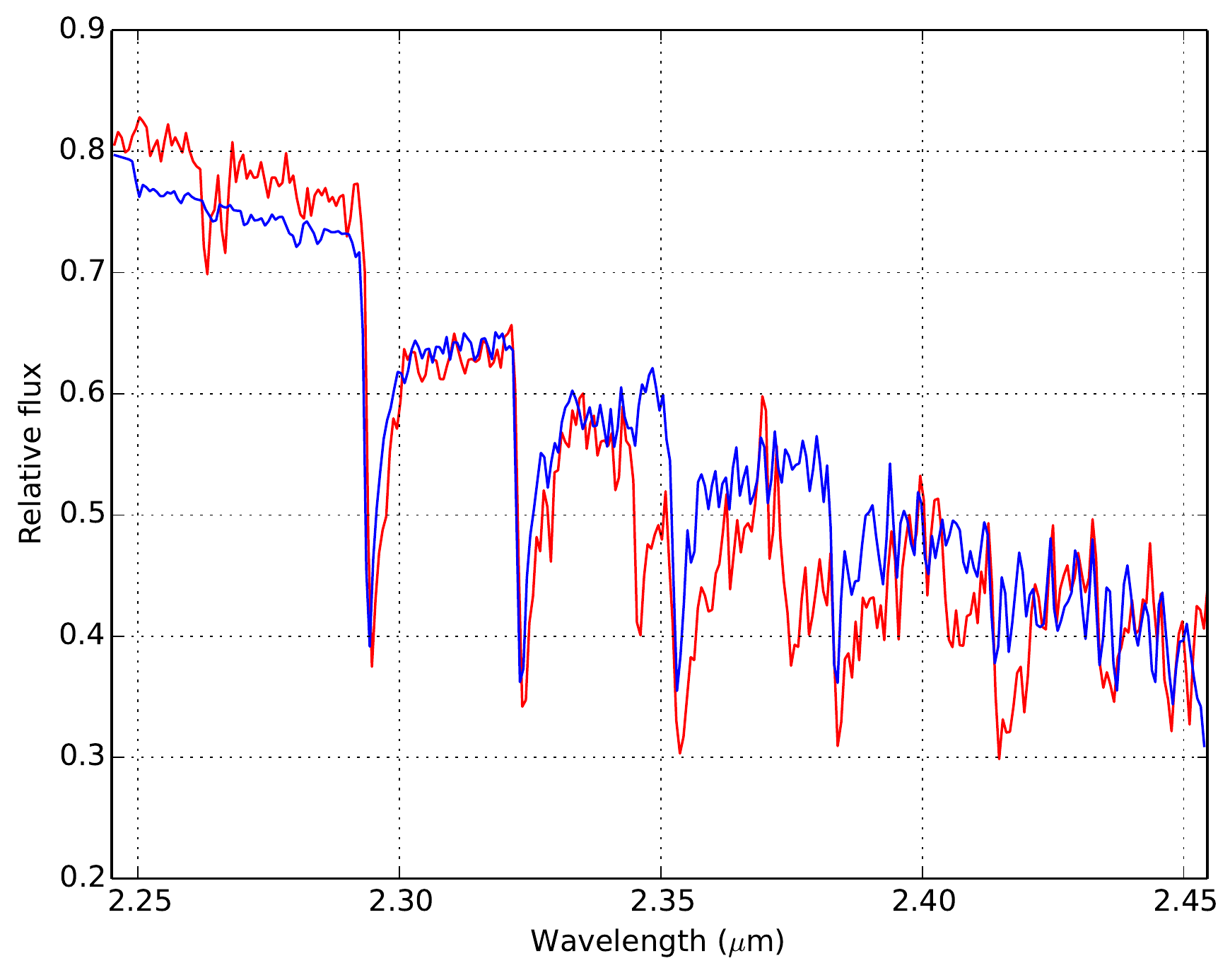}}
		\caption{The black line is the spectrum obtained from the Betelgeuse AMBER data and the red line is the spectrum obtained from the single layer model. We used the best fit values from Table \ref{Eq_Results}.}
		\label{Fig_ModelSpec}
	\end{figure}
	
	The best fit parameters are presented in Table \ref{Eq_Results}. These results are robust enough to be insensitive to a noise of 10\% of the observed data. It is noteworthy that the column densities converged for the same $(T_\mathrm{MOL},\theta_\mathrm{MOL})$ values: this confirms our hypothesis of a single molecular layer containing both CO and H$_2$O.

	\begin{table}[!ht]
		\caption{Best fit values with our single thin layer model of the MOLsphere.}             
		\label{Eq_Results}      
		\centering          
		\begin{tabular}{l l} 
		\hline\hline
		Parameter & Value \\
		\hline
		\noalign{\smallskip}
		$\theta_\star$ & 41.01~mas (fixed)\\
		T$_\mathrm{MOL}$ & $2300 \pm 120$~K\\
		$\theta_\mathrm{MOL}$ & $51.38 \pm 1.71$~mas\smallskip\\
		N$_\mathrm{CO}$ & $3.01^{+2.0}_{-0.5} \times 10^{21} ~\mathrm{cm}^{-2}$\smallskip\\
		N$_{\mathrm{H}_2\mathrm{O}}$ & $3.28^{+1.7}_{-0.5} \times 10^{20} ~\mathrm{cm}^{-2}$\\
		$\chi^2_\mathrm{red}$ & $\sim 6$\\
		\hline
		\end{tabular}
		\tablefoot{The error bars were computed by solving the equation $\chi^2_\mathrm{red}$(T$_\mathrm{MOL}$, $\theta_\mathrm{MOL}$, N$_\mathrm{CO}$, N$_{\mathrm{H}_2\mathrm{O}}$) = 2$\chi^2_\mathrm{red, min}$).}
	\end{table} 
	
	Our best fit only matches the spectral domain around the band heads of the first two CO overtone lines ($2.245 < \lambda < 2.348~\mu$m, see Fig. \ref{Fig_ModelSpec}). Therefore, we deduced that the absorption lines for $\lambda > 2.348~\mu$m cannot be reproduced by considering material characterized by the best fit parameters derived for our single layer model. In other words, to successfully model these absorption features, we would need to introduce at least another layer in the model, which would be located at a different distance from the photosphere. This issue is adressed in Sect. \ref{Sect_Discussion}.
	
	\onlfig{
	\begin{figure*}[!ht]
			\centering
			\includegraphics[width=8cm]{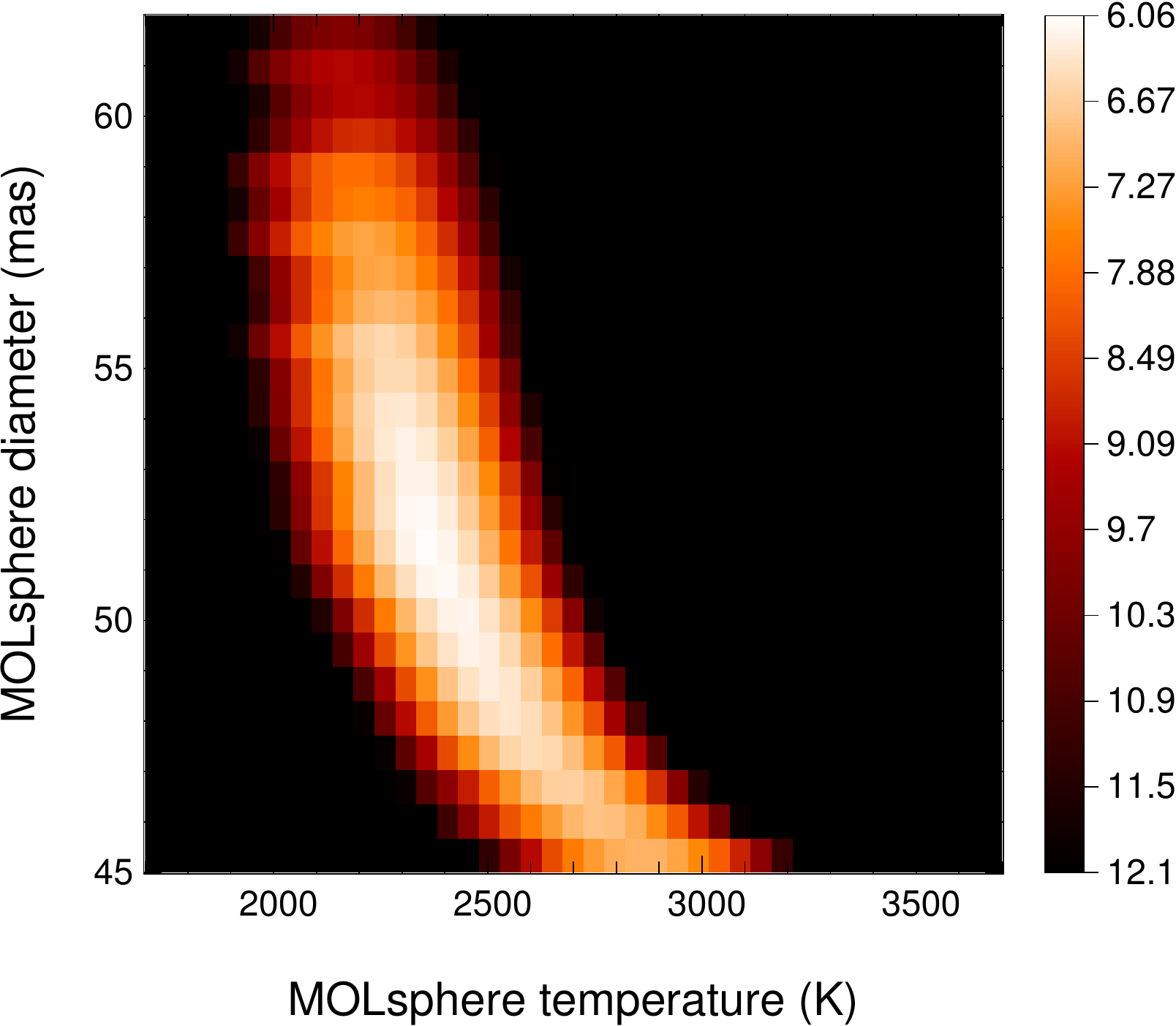}
			\quad
			\includegraphics[width=8cm]{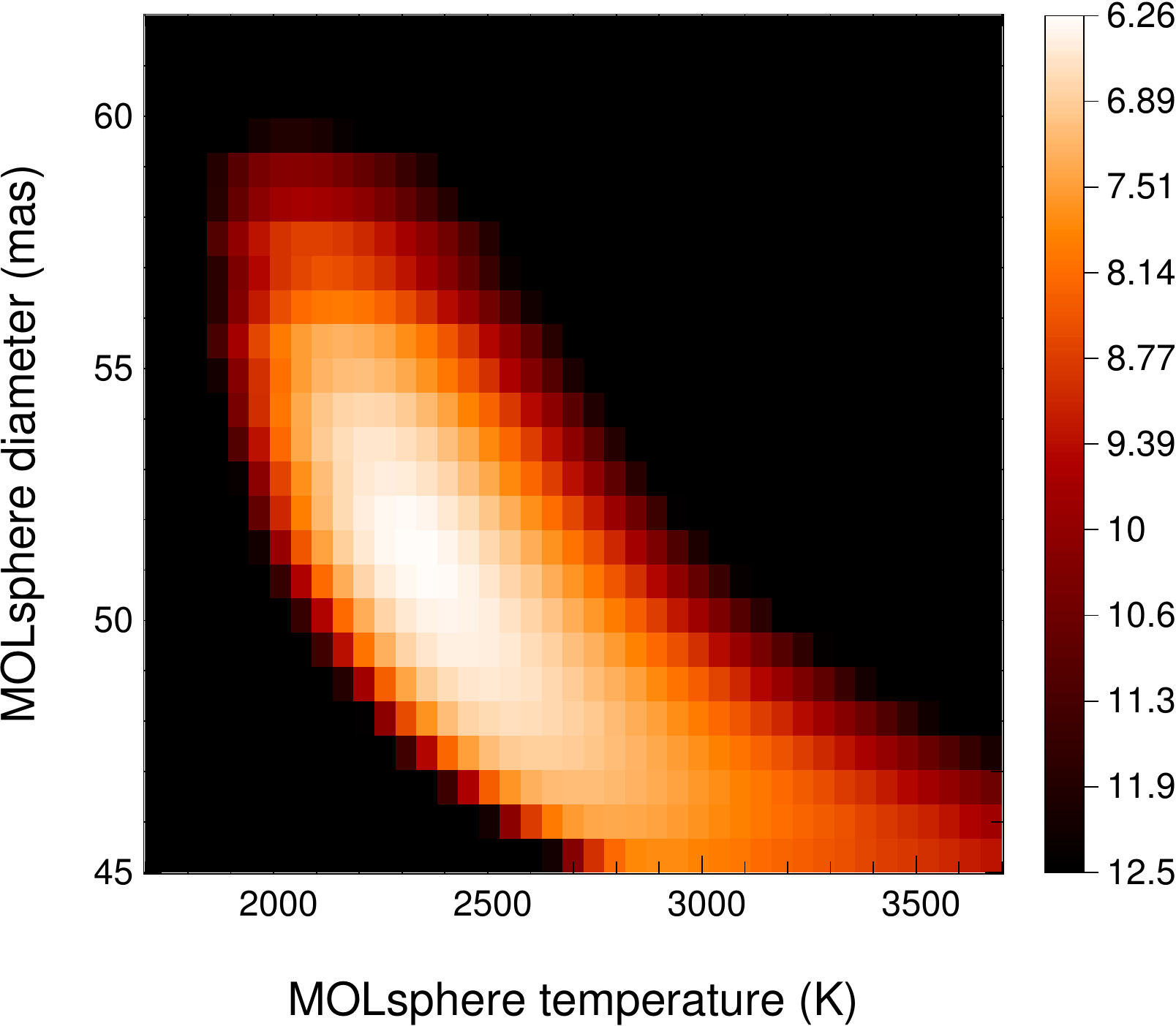}
			\caption{$\chi^2$ map of the single layer model. \textit{Left: }The CO column density is fit on each cell of the grid for a constant $N_{{\mathrm{H}_2\mathrm{O}}} = 3.28 \times 10^{20} \, \mathrm{cm}^{-2}$. \textit{Right: }The H$_2$O column density is fit on each cell of the grid for a constant $N_{\mathrm{CO}} = 1.53 \times 10^{21} \, \mathrm{cm}^{-2}$.}
			\label{Fig_Chi2Map}
	\end{figure*}
	}	
	
	\subsubsection{Photocenter position}\label{SubSect_DP}
	
	\begin{figure}[!ht]
		\centering
		\resizebox{\hsize}{!}{\includegraphics{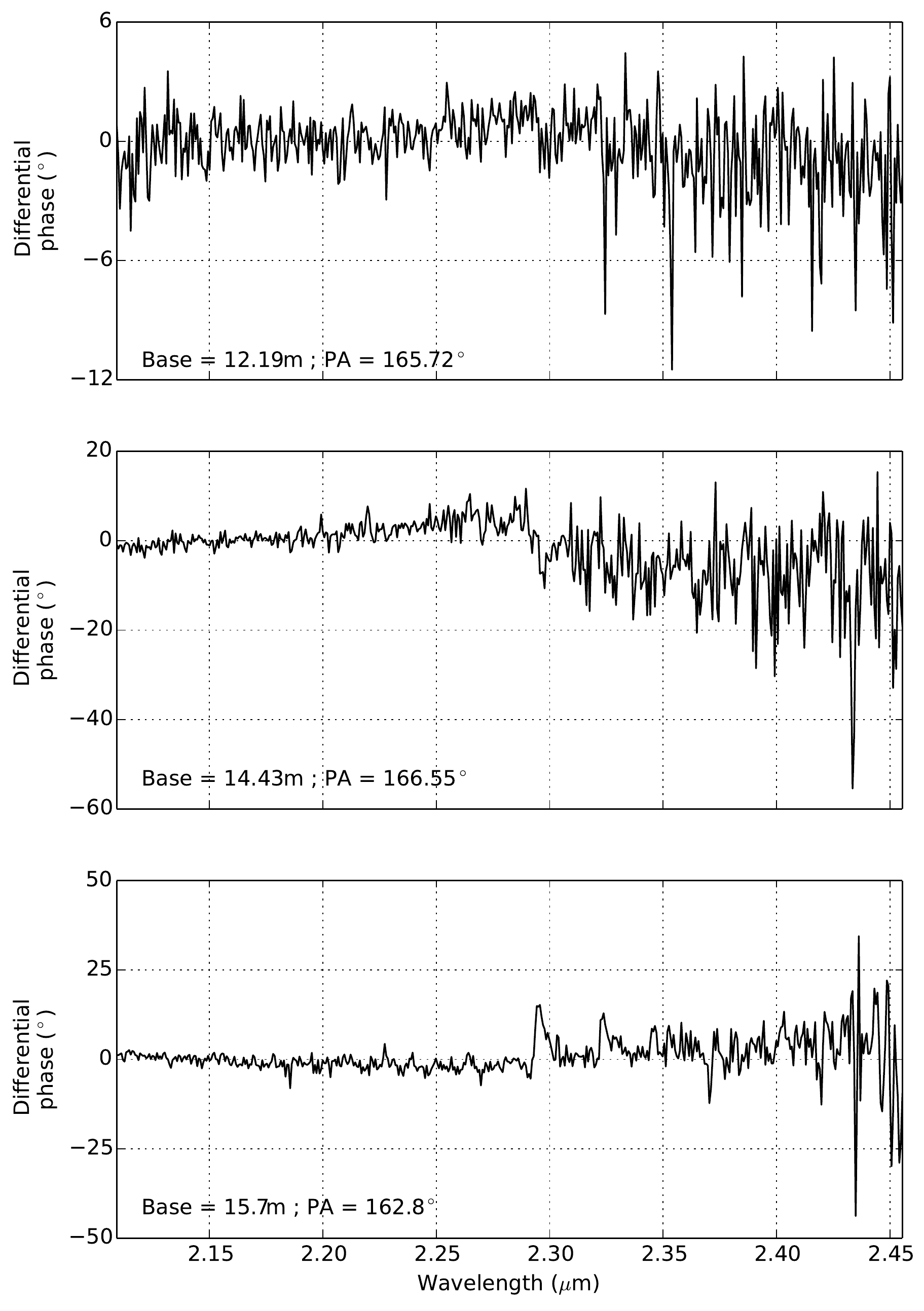}}
		\caption{Differential phases of our datasets below 55 arcsec$^{-1}$ (sampling the first and second lobe of the visiblity function). We selected measurements not sampling a node of the visibility function.}
		\label{Fig_DP}
	\end{figure}
	
	Figure \ref{Fig_DP} represents the DPs in the first and second lobe (spatial frequencies below 55 arcsec$^{-1}$), which correspond to the visibilities we fitted in Sects. \ref{SubSect_Continuum} and \ref{SubSubSect_MOLsphere}. We selected the datasets not sampling a node of the visibility function to avoid phase oscillations between 0 and $\pi$. Therefore, the non-zero and non-$\pi$ DP values in the absorption lines of the spectral range indicate a displacement of the photocenter between the continuum and the CO and H$_2$O absorption domain, which are observations already made and modeled by \citetads{2009A&A...503..183O,2011A&A...529A.163O}. Unfortunately, our medium spectral resolution prevents us from performing the same analysis due to contamination by adjacent spectral channel, which would bias the DP in the lines.
	
	\subsection{H band}
	
	Data in the H band were successfully reduced and calibrated, but many absorption features are present at those wavelengths, preventing us from isolating the continuum to perform our UD and LDD fits. We tried to use our single thin layer model but attempts with parameters around the best fit values found in Sect.~\ref{SubSubSect_MOLsphere} gave an inconsistent spectrum. From our several trials, we think that more molecules or more layers are required to account for all the absorption features of the H band. 

\section{Numerical simulations: Deviations from central symmetry} \label{Sect_Simulations}

	We saw in Sect. \ref{SubSect_Continuum} that the UD and LDD models poorly reproduce the spatial frequencies higher than 60~arcsec$^{-1}$. We investigate this by using a radiative hydrodynamics (RHD) simulation that is obtained with the CO$^5$BOLD code \citepads{2012JCoPh.231..919F} to unveil the signature of convection patterns on the star photosphere. We used the non-gray model st35gm03n13, which is described in detail in \citetads{2011A&A...535A..22C}. The grid resolution is $235^3$ points with a step of $8.6~R_\odot$. The parameters of the star used in the model are presented in Table \ref{Table_Simu_Param} with the corresponding values for Betelgeuse.
	
	\begin{table}[h!]
		\caption{Characteristics of Betelgeuse vs the parameters of the model of the RHD simulation.}             
		\label{Table_Simu_Param}      
		\centering          
		\begin{tabular}{l l l} 
			\hline\hline
			Parameter & Betelgeuse & Model \\
			\hline
			\noalign{\smallskip}
			M (M$_\odot$) & 11.6 \citepads{2011ASPC..451..117N} & 12.0\\
			L (L$_\odot$) & $1.27 \times 10^5$ (this work) & $8.95 \times 0^4$ \\
			R (R$_\odot$) & 897 (this work) & $846$\\
			T$_\mathrm{eff}$ (K) & 3640 \citepads{2005ApJ...628..973L} & $3430$\\
			$\log(g)$ & $-0.300$ \citepads{2008AJ....135.1430H} & $-0.354$\\
			\hline
		\end{tabular}
	\end{table}
	
	Several snapshots of the simulation were extracted at various evolution times, thus representing different realizations of the model. The intensity maps of these snapshots were computed using 3D pure-LTE radiative transfer with the OPTIM3D code \citepads{2009A&A...506.1351C}, and each snapshot was rotated around its center with a step of 10$^\circ$, as we do not know the real orientation of the model relative to the star on sky. We obtained a grid of rotated snapshots, each one becoming a realization of the convective pattern.
	
\begin{figure}[!ht]
		\centering
		\resizebox{\hsize}{!}{\includegraphics{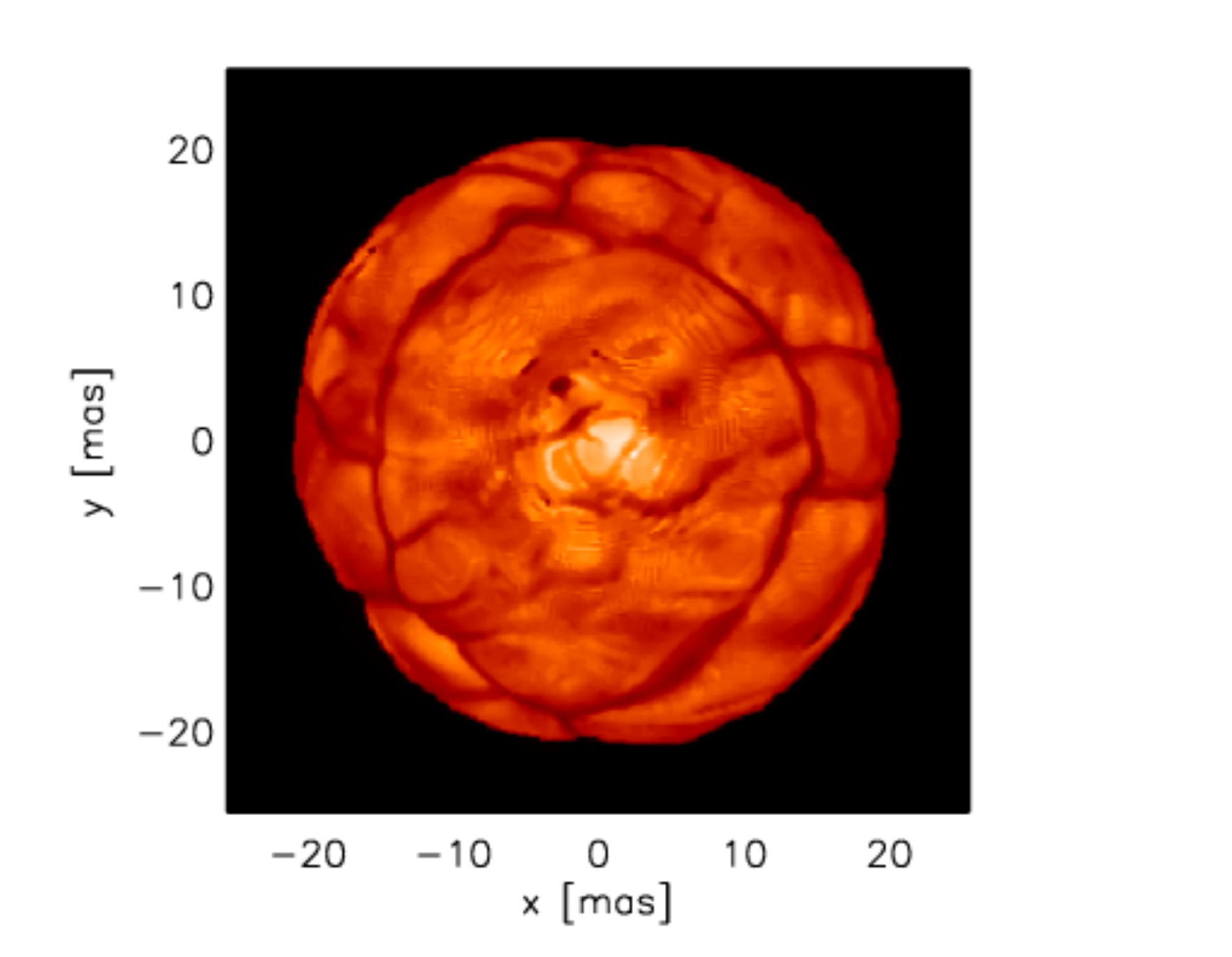}}
		\resizebox{\hsize}{!}{\includegraphics{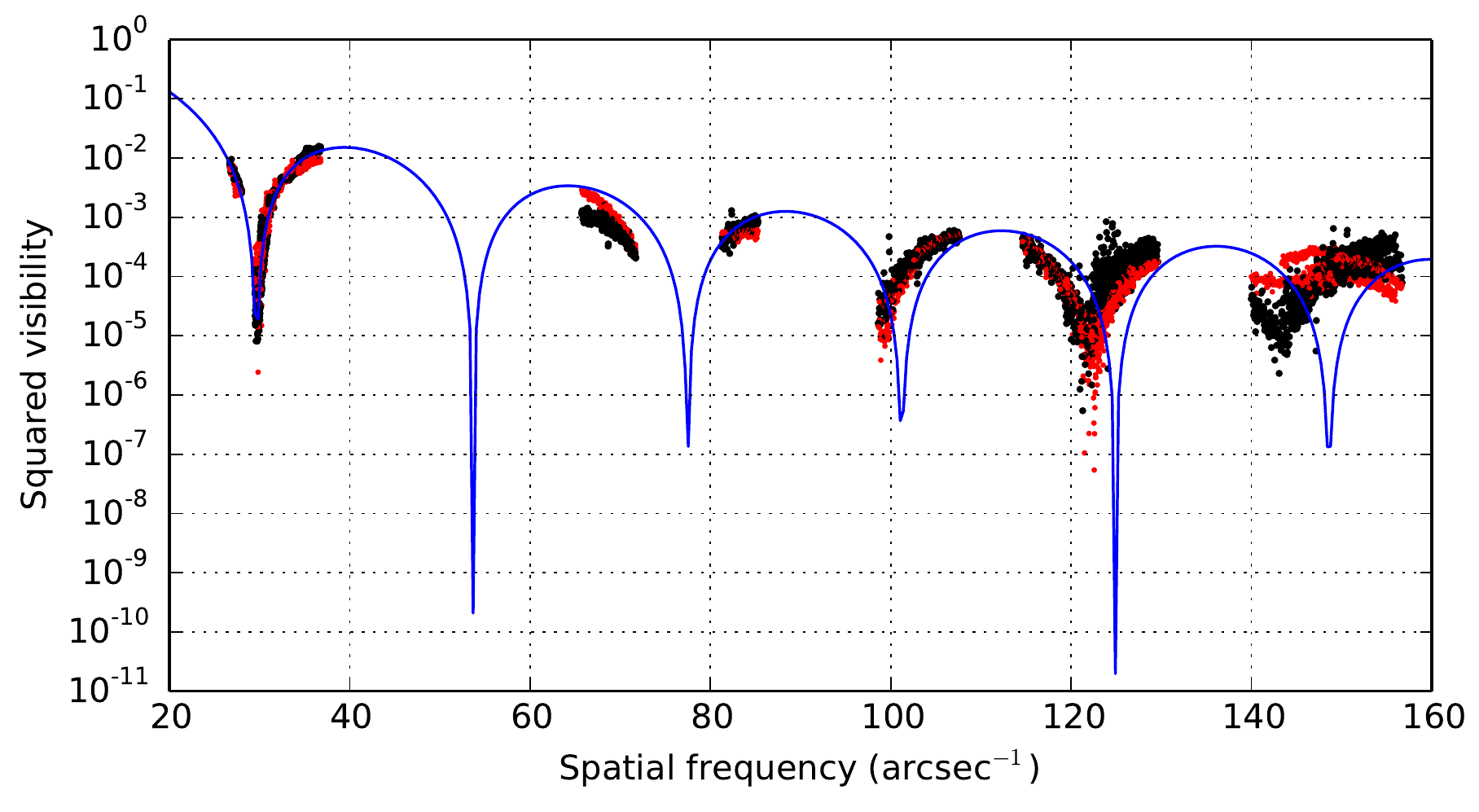}}
		\caption{\textit{Top:} Intensity map of the best-matching snapshot of RHD simulation at 2.2 $\mu$m (linear scale with a range of [0; 130000] erg.cm$^{-2}$.s$^{-1}$.\AA$^{-1}$. \textit{Bottom:} Comparison of the best fit snapshot with the AMBER continuum data. The AMBER squared visibilities are in black and the best fit squared visibilities of the simulation in red. The best fit LDD model is represented by the blue continuous line.}
		\label{Fig_VisSimu}
	\end{figure}

	We computed some intensity maps from the simulation and for wavelengths corresponding to the AMBER observations in the continuum region ($\lambda < 2.22 \mu$m). Then we derived the visibilities corresponding to the (u,v) sampling of our AMBER dataset. The method used to extract the interferometric observables is described in \citetads{2010A&A...515A..12C}. We computed the $\chi^2$ for each realization of the convective pattern of our grid. This allowed us to select the best fit snapshot and rotation angle for which we obtained $\chi^2_r = 7.47$. This best fit model is illustrated on Fig. \ref{Fig_VisSimu}.
	
	When we consider all our data and not only the first and second lobes of the visibility function, our best fit LDD presented in Sect. \ref{SubSect_Continuum} gives $\chi^2_\mathrm{r} = 46.4$. Therefore, with this best fit snapshot, we manage to reproduce the shape of the high spatial frequencies signal better than the LDD model. This is another piece of evidence of the convection on the photosphere of Betelgeuse, which was already revealed with the interpretation of interferometric observations from the optical to the infrared domains \citepads{2009A&A...506.1351C,2010A&A...515A..12C}.
	
	However, the visibilities are not perfectly reproduced, as the minimum reduced $\chi^2_r$ is greater than 1. Several reasons can explain this: the most obvious is that the snapshot is not reproducing the visibilities, because its intensity distribution differs from the photosphere of the star. However, one should also consider that there may still be partially biased datasets, even if we discard most of them (see Sect. \ref{SubSect_Reduction}). This is particularly true for the low visibilities encountered in the higher order lobes: the largest FINITO tracking losses were discarded but the remaining dataset is probably not entirely clean, and we do not have any way to characterize it. Moreover, the limited (u,v) coverage of this three-telescope dataset (Fig. \ref{Fig_uv_cov}) cannot fully describe the whole convective pattern of Betelgeuse. Trying to strongly constrain the model with these observations would be irrelevant; therefore, we did not expand our sample of snapshots to lower the $\chi^2$.
	
	\begin{figure}[!ht]
		\centering
		\resizebox{\hsize}{!}{\includegraphics{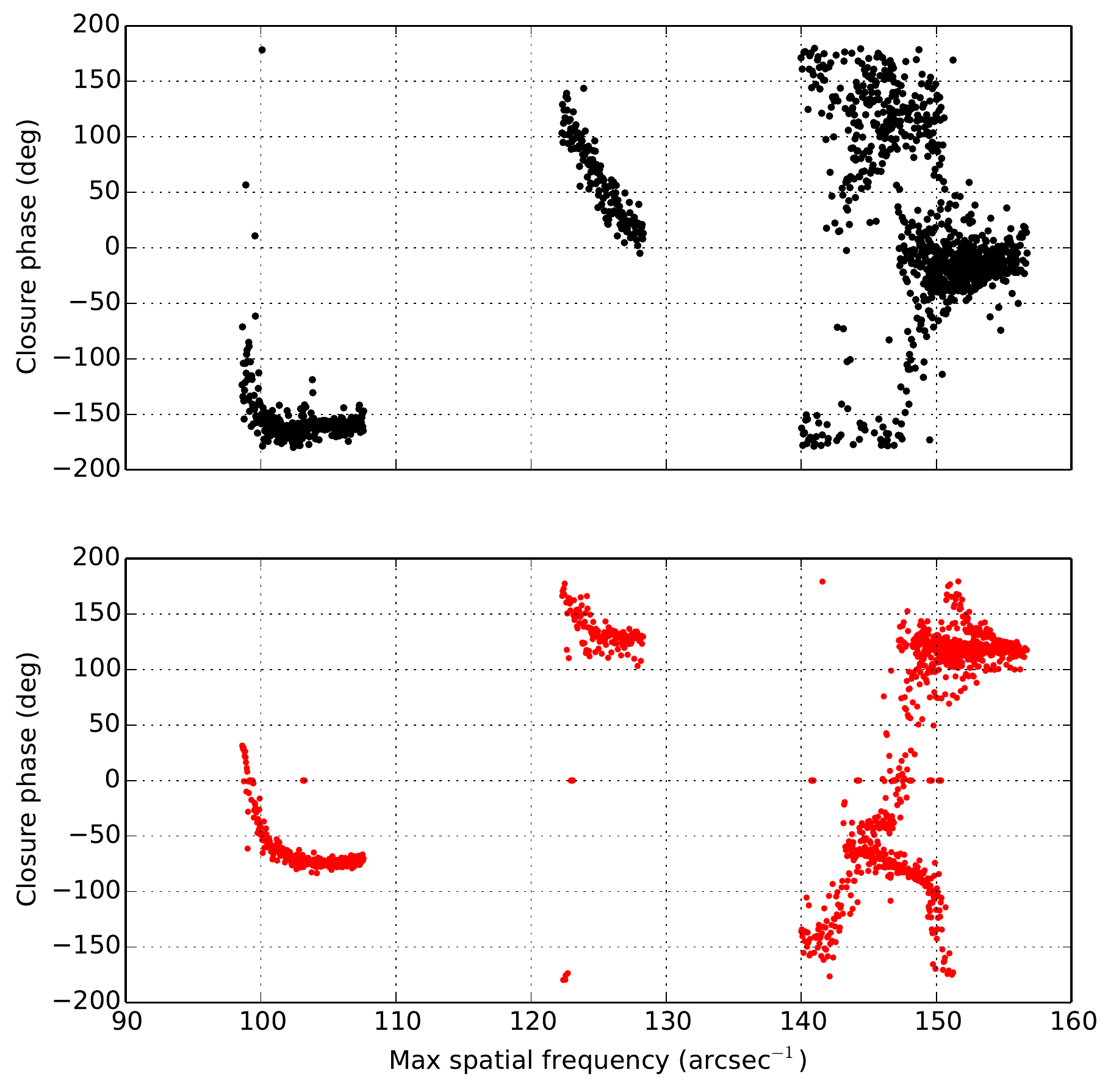}}
		\caption{Comparison of the closure phases of the best fit snapshot (bottom) with the AMBER continuum closure phases (top).}
		\label{Fig_CPSimu}
	\end{figure}
	
	Figure \ref{Fig_CPSimu} represents the closure phases of both the continuum data and the best fit model. This quantity is strongly sensitive to asymmetries on the observed target; thus, it is directly related to a particular realization of the convection pattern, in the case of a RSG. Consequently, in contrast to the visibilities, it is as affected by the contrast and the size of the cells as it is by their position on the disk. Therefore, it is difficult for a given snapshot to reproduce its shape, even by dramatically increasing the statistics (and the computation time). For this reason, we only fitted the squared visibilities. However, it is interesting to remark that the general shape of the closure phase is qualitatively well reproduced below 130~arcsec$^{-1}$, even if the absolute value does not correspond. It is a strong indication that the closure phases agree with a convective pattern.
	
\section{Discussion} \label{Sect_Discussion}
	
	The results (Table \ref{Eq_Results}) from our model fit described in Sect. \ref{SubSubSect_MOLsphere} do note agree with the parameters of the outer molecular layer as described in \citetads{2009A&A...503..183O} and \citetads{2011A&A...529A.163O}. Our inability to reproduce the observed spectrum of Betelgeuse in the absorption lines of CO and H$_2$O for $\lambda > 2.348~\mu$m suggests that our model is incomplete. We think that at least two molecular layers are requested to model the molecular material close to Betelgeuse (a first step towards an extended layer model), which was the model already developed by these authors to analyze their high spectral resolution AMBER observations. They computed the CO column densities separately for both layers, using a photospheric model with \citetads{2006ApJ...645.1448T} parameters for the inner layer. Similar results were obtained by \citetads{2013A&A...555A..24O} with their observations of \object{$\alpha$ Sco}. It is interesting to notice the remarkable similarity in the observable characteristics of those two stars: the same model of a two layer MOLsphere is giving similar values for the molecule column densities. We tried to fit our data using this two-layer model but the inner layer of the MOLsphere was converging to the photosphere and increasing the $\chi^2$ (by a factor from three to four). Adding a second layer also adds four parameters (the layer angular diameter, its temperature, and the two column densities). The problem becomes highly degenerated as both layers contributes to the whole absorption domain. We could use previous measurements to initially constain the model but this would lead to strong bias, particularly if the material injection in the MOLsphere is indeed episodic. Moreover, it appears that we could not use this two-layer model in our case, because of the lack of completely resolved spectral features in our spectrum with the medium spectral resolution. Our parameters also differ from those proposed by \citetads{2004A&A...418..675P,2007A&A...474..599P}. This is not surprising as the authors of these two last papers used different wavelength ranges in their analysis: we may not be observing the same region of the MOLsphere. Actually, these different points are part of a wider discussion on the hypothesis of the MOLsphere: instead of one or several thin layers, we may be dealing with a thick layer with a spatially inhomogeneous (but continuous) composition in the radial direction. This envelope is probably not at LTE, and on-going studies are already exploring this path \citepads{2012MNRAS.427...27B,2013ApJ...764..115B,2013EAS....60..111L}. Finally, the envelope photocenter may be offset with respect to the center of the stellar disk, as suggested by the differential phases plotted in Fig. \ref{Fig_DP}. Therefore, observations with higher spectral resolution and also a better (u,v) coverage are needed to get a complete overview of the MOLsphere as well as more physically realistic models to match the interferometric observations of the close envelope.
	
	Our model of a single thin molecular layer (Sect. \ref{SubSubSect_MOLsphere}) allows us to derive the abundances of CO and water vapor around Betelgeuse (Table \ref{Table_Molecules}) by considering that Betelgeuse is $197 \pm 45$~pc away \citepads{2008AJ....135.1430H}. The AMBER field-of-view is estimated to be 300~mas \citepads{2010A&A...520L...2A}, but \citetads{2012AJ....144...36O} detected CO up to several arcsec away from the star using the Combined Array for Research in Millimeter-wave Astronomy (CARMA). Thus, the total molecular mass observed with AMBER in the K band does not correspond to the whole envelope of Betelgeuse but may be compared to the estimated mass loss of the star, 2 -- 4 $\times 10^{-6}$~M$_\odot$.yr$^{-1}$ (\citeads{1990ApJS...73..769J} and \citeads{2013EAS....60..307V}). Considering both CO and H$_2$O, the material observed with AMBER corresponds to $5.46^{+3.4}_{-1.0} \times 10^{-6}$~M$_\odot \sim 1.8$~yr of mass loss. Therefore, the mass of the molecular material we observe matches roughly the yearly mass loss of Betelgeuse. Considering an oxygen abundance around Betelgeuse of $\log \epsilon(O) = 8.8$ \citepads{1984ApJ...284..223L}, we derive a value of M$_\mathrm{O} \sim 10^{-4}$ M$_\mathrm{H}$ in the envelope of the star, meaning that oxygen-bearing molecules should represent a tiny fraction of the total CSE  in mass, yet we observe that it already corresponds to more than the yearly material expelled from the star in the region reached in one year around it. This paradox can be explained in several ways. First, as discussed by \citetads{2013EAS....60.....K}, the mass loss could be episodic; therefore, if $\alpha$ Ori went through a high mass loss episode recently, one expects to observe a high density of material in its close environment. Such an event could be driven by the convection, as suggested by \citetads{2007A&A...469..671J}. The asymmetric and inhomogeneous structures observed by \citetads{2009A&A...504..115K,2011A&A...531A.117K} could also correspond to such a transitional events. One also has to take into account that not all the material injected in the MOLsphere effectively participates in the mass loss: \citetads{2011A&A...529A.163O} observed both upward and downward motions in the CO MOLsphere, indicating that part of the molecules are falling back on the star. Therefore, observing a higher density of material in the CO and H$_2$O MOLsphere than what is predicted, when only the yearly mass loss rate is considered, is consistent, but more observations are needed to distinguish between those different scenarios. One can also note that these two explanations do not exclude each other, and it is likely that both contribute to the high molecular density observed in the MOLsphere.
	
	\begin{table}[h!]
		\caption{Total mass of CO and H$_2$O around Betelgeuse, as derived from our single thin layer model.}             
		\label{Table_Molecules}      
		\centering          
		\begin{tabular}{l l} 
			\hline\hline
			Parameter & Value\\
			\hline
			\noalign{\smallskip}
			m$_\mathrm{CO}$ & $5.1^{+3.4}_{-0.9} \times 10^{-6}$~M$_\odot$\smallskip\\ 
			m$_{\mathrm{H}_2\mathrm{O}}$ & $3.6^{+1.9}_{-0.5} \times 10^{-7}$~M$_\odot$\smallskip\\
			\hline
		\end{tabular}
	\end{table}

\section{Conclusion} \label{Sect_Conclusion}

	We obtained a new measurement of the angular diameter of Betelgeuse in the K band by considering only the continuum. This measurement, along with the previous values collected during almost twenty years, does not indicate any monotone variation of the star diameter, unlike the 11~$\mu$m observations of \citetads{2009ApJ...697L.127T}. As \citetads{2011ASPC..448.1025R} and \citetads{2011A&A...529A.163O} already explained, the 11~$\mu$m measurements are more sensitive to the evolution of the molecular and dusty layers than the K-band observations, which mostly probe the star, particularly in the continuum.

	We spatially resolved Betelgeuse in water vapor and carbon monoxide in the K band, using the medium spectral resolution of VLTI/AMBER. We obtained values for the column densities of a model that consists of a single thin layer for the MOLsphere and Kurucz model for the star.
	
	Finally, using a RHD simulation, we bring new evidence for a convection pattern on Betelgeuse's surface. The sample of simulation snapshots reproduces the shape of the visibility signal of our AMBER dataset in the continuum domain at high spatial frequencies.

\begin{acknowledgements}
	We are grateful to ESO’s Director General Prof. Tim de Zeeuw for the allocation of observing time to our program, as well as to the Paranal Observatory team for the successful execution of the observations.
	The research leading to these results has received funding from the European Community's Seventh Framework Programme under Grant Agreement 226604.
	This research received the support of PHASE, the high angular resolution partnership between ONERA, Observatoire de Paris, CNRS and University Denis Diderot Paris 7.
	We acknowledge financial support from the “Programme National de Physique Stellaire” (PNPS) of CNRS/INSU, France. 
%       STR acknowledges partial support by NASA grant NNH09AK731. 
	We used the SIMBAD and VIZIER databases at the CDS, Strasbourg (France), and NASA’s Astrophysics Data System Bibliographic Services. 
	This research has made use of the  \texttt{AMBER data reduction package} of the Jean-Marie Mariotti Center\footnote{Available at \url{http://www.jmmc.fr/amberdrs}}
	We thank the referee, Dr. Bernhard Aringer, for his suggestions that led to improvements of this article.
	
\end{acknowledgements}

\bibliographystyle{aa}
\bibliography{./biblio.bib}

\Online

%\listofobjects

\end{document}